\documentclass[11pt,a4paper]{article}

\def\beq{\begin{equation}}
\def\eeq{\end{equation}}
\def\beqa{\begin{eqnarray}}
\def\eeqa{\end{eqnarray}}

\usepackage{jheppub}
\title{Observer dependence of bubble nucleation and Schwinger pair production}

\author[a,b]{Jaume Garriga,}
\author[b]{Sugumi Kanno,}
\author[c]{Misao Sasaki,}
\author[d]{Jiro Soda,}
\author[b]{Alexander Vilenkin}
\affiliation[a]{Departament de F{\'\i}sica Fonamental i \\
Institut de Ci{\`e}ncies del Cosmos, 
Universitat de Barcelona,\\
Mart{\'\i}\ i Franqu{\`e}s 1, 08028 Barcelona, Spain}
\affiliation[b]{Institute of Cosmology, Department of Physics and Astronomy, 
Tufts University, Medford, Massachusetts 02155, USA}
\affiliation[c]{ Yukawa Institute for Theoretical Physics, Kyoto University,
Kyoto 606-8502, Japan}
\affiliation[d]{
Department of Physics, Kyoto University, Kyoto 606-8502, Japan}

\abstract{
Pair production in a constant electric field is closely analogous to
bubble nucleation in a false vacuum.  The classical trajectories of the 
pairs are Lorentz invariant, but it appears that this invariance should be
broken by the nucleation process. Here, we use a model detector, consisting of other particles 
interacting with the pairs, to investigate how pair production is seen by different Lorentzian observers.
We focus on the idealized situation where a constant external electric field is present for an infinitely long time, and we
consider the in-vacuum state for a charged scalar field that describes the nucleating pairs. 
The in-vacuum is defined in terms of modes which are positive frequency in the remote past.
Even though the construction uses a particular reference frame and a gauge where the vector
potential is time dependent, we show explicitly that the resulting quantum state is Lorentz invariant.
We then introduce a ``detector" particle which interacts with the nucleated pairs, 
and show that all Lorentzian observers will see the particles and antiparticles
nucleating preferentially at rest in the detector's rest frame.  Similar conclusions are
expected to apply to bubble nucleation in a sufficiently long lived vacuum.
We also comment on certain unphysical aspects of the Lorentz invariant in-vacuum, associated
with the fact that it contains an infinite density of particles. This can
be easily remedied by considering Lorentz breaking initial conditions.

}

\keywords{Schwinger pair creation, bubble nucleation, Lorentz violation}
\preprint{YITP-12-68, KUNS-2410}

\begin{document}
\maketitle

\section{Introduction}

Vacuum decay through bubble nucleation was first analyzed in the 
ground-breaking paper by Voloshin, Kobzarev and Okun (VKO) \cite{VKO}.  
Their analysis also uncovered an unexpected problem.  In the semiclassical
 picture, the bubble is formed momentarily at rest; then it expands, 
rapidly approaching the speed of light.  Lorentz invariance of the vacuum 
requires that the probability of nucleating a bubble should be the same 
in all inertial frames.  This suggests that the rate of bubble formation 
per unit spacetime volume should include an integral over the rest frame
 of nucleation, that is, over the Lorentz group.  This integral is 
of course divergent, so one obtains a meaningless infinite answer
\footnote{For a recent discussion of related issues, see \cite{Dvali,GSV,GV12}.}.

An elegant resolution of the paradox was given by Coleman \cite{Coleman}, 
who developed an instanton method for calculating the bubble nucleation rate. 
The instanton in this case is $O(4)$ invariant, and its Lorentzian continuation
is invariant with respect to Lorentz boosts. From this Coleman concluded 
that ``an expanding bubble looks the same to all Lorentz observers, and
to integrate over the Lorentz group is to erroneously count the same 
final state many times". 

There is little doubt that Coleman's calculation of the nucleation rate
is correct.  However, some intriguing questions about the bubble nucleation
process remain unanswered.  The Lorentzian continuation of the instanton
solution describes a bubble which contracts from infinite size, bounces 
at a minimum radius, and then re-expands.  This solution is Lorentz invariant,
but it appears that the contracting part of it is unphysical and needs to be cut off. 
However, the cutoff would necessarily break the Lorentz symmetry, and we would 
be back to the VKO problem.  The answer that is usually given is that 
the semiclassical approximation breaks down near the nucleation point 
(that is, near the bounce at the minimal radius). So there is no well 
defined cutoff surface.  The bubble emerges from a fuzzy quantum region,
 and only its late-time asymptotic behavior has classical meaning.

This may be true, but we can still ask how different observers would see 
the nucleation process.  An observer can fill space with detectors measuring
 the scalar field of the bubble.  This would yield a complete spacetime 
history of the nucleation process.  In particular, the rest frame of bubble 
nucleation may be determined in this way.

One can anticipate several possible scenarios that can emerge from such an 
analysis.  (A) It could be that the frame of nucleation simply coincides
 with the rest frame of the detectors.  In other words, each observer will 
see bubbles forming at rest in her own rest frame.  (B) It is conceivable that the contracting part of the bubble history 
is not completely cut off and can be at least partially observed.  In fact, in the case of scenario (A), an observer 
moving relative to the detectors will see the detection begin on the contracting part of the bubble trajectory.  (C) Yet another possibility 
is that the frame of nucleation is influenced by how the decaying false 
vacuum was set up.  If the false vacuum has zero energy, its spacetime is 
flat, and the space will be filled by nucleating bubbles in a finite amount 
of time.  This implies that the false vacuum could not have existed forever;
 it must have been created in some manner in the past.  We could imagine, 
for example, that the false vacuum was set up on some spacelike surface, 
e.g., on a hyperplane $t={\rm const}$.  The preferred frame of reference
 specified by this choice could in principle determine the frame of bubble
 nucleation.

If the false vacuum has positive energy, it is described by de Sitter space,
 and the bubble nucleation process can continue forever.  However, the geometry
 of the full de Sitter space is similar to that of the worldsheet of a bubble:
 it describes a universe contracting from infinite size, bouncing at a minimum
 radius and re-expanding.  Bubbles would fill the universe during the 
contracting phase, so one needs to introduce a cutoff.  This would break 
the de Sitter symmetry and define a preferred frame\footnote{The conclusion 
that the false vacuum state cannot be extended to the infinite past has 
been shown to hold in the most general case, without assuming homogeneity
 or isotropy \cite{Borde,BGV}.}.
One might think that the memory of the preferred frame would gradually 
fade away and would have no effect on bubbles nucleating at late times. 
 However, it was shown in \cite{GGV} that a peculiar ``persistence of memory"
 effect does not fade away and persists at arbitrarily late times. 
 Thus the possibility (C) that the frame of nucleation is influenced 
by the initial conditions remains a viable option.

In the present paper we shall attempt to address these issues using pair
 production in a constant electric field as a model for bubble nucleation.
  We shall consider charged spin-0 particles, described by a complex scalar
 field $\phi(t,x)$, in a constant electric field $E$ in a $(1+1)$-dimensional
 Minkowski space.  Such an electric field is Lorentz invariant, and pair
 production by the Schwinger process \cite{Schwinger:1951nm} is closely analogous to bubble nucleation.
  It is conceivable that despite the invariance of the background, the 
Lorentz symmetry could be spontaneously broken by the quantum state of 
the $\phi$-field.  Indeed, particles and antiparticles of the pairs are 
accelerated by the electric field in opposite directions, and since the 
number of the pairs is constantly growing, one can expect a growing 
expectation value of the electric current $\langle J^1\rangle$. 
 Any non-zero value of $\langle J^\mu \rangle$ would break the Lorentz 
invariance.  However, the situation is complicated by the fact that 
the current $\langle J^\mu \rangle$ is formally infinite and needs to 
be renormalized.

Hence, the first part of our project is to investigate the invariance 
properties of the in-vacuum state, defined by requiring that the mode
 functions of the $\phi$-field are positive-frequency at $t\to -\infty$. 
 This quantum state has been extensively discussed in the literature 
(see, e.g., \cite{Grib,Brout:1995rd,Massar,Gabriel,Tanji:2008ku} and references therein), but to
 our knowledge its invariance properties have not yet been studied.
We show by a direct calculation that the in-vacuum state {\it is} Lorentz invariant.

Nonetheless, as we shall see, divergent terms in the expectation values of physical observables, like the current 
$J^\mu$ or the energy-momentum tensor $T_{\mu\nu}$, 
do not seem to admit any Lorentz invariant regulators.   
The unusual properties of the in-vacuum state are due to the fact that at any finite time it contains an infinite number of out-particles. 
In a more realistic model, the electric field 
would be cancelled by the back-reaction of the created 
pairs, and only a finite density of particles would be produced. Alternatively, in the context where the electric field is external, we can still consider states where the 
number density of pairs is finite at any given moment of time. For instance, we could choose initial conditions such that the number density of pairs vanishes on some 
initial surface $t={\rm const}.$\footnote{ The number density of pairs is not a sharply defined quantity. Nonetheless, it can be defined by using the instantaneous
Hamiltonian diagonalization.  Such states (with vanishing particle number at $t = 0$) have been discussed in Refs. \cite{Grib} and \cite{Tanji:2008ku}. 
The initial number of particles could also be rigorously defined by considering an electric field which vanishes at $t\to -\infty$ and then
turned on at some time in the past.  An abrupt turn-on at an instant of time was considered in Refs. \cite{Tanji:2008ku, Gavrilov:1996pz, Gavrilov:2012jk}, and an adiabatic turn-on with $E(t) \propto 1/\cosh^2 (t/T)$ was discussed in Ref. \cite{Grib}.
}
We shall come back to this issue in Section 4. 

Despite its somewhat unphysical properties, we find that the in-vacuum is a useful laboratory for studying the bubble 
nucleation process, at least in the limit where the initial conditions are removed sufficiently far in the past.  
In this paper we shall consider tree-level interactions between the pairs and the detector. The kinematics of these interactions
is such that, out of the infinite bath of created particles, only those whose momentum relative to the detector is in a certain range will 
have a chance to interact with it. In this sense, most particles in the bath are invisible, and their infinite density is irrelevant.

To be specific, we model the detector by introducing two additional scalar fields: 
a charged scalar $\psi(t,x)$ and a neutral scalar $\chi(t,x)$, with the 
interaction Lagrangian
\beq
L_{\rm int}=-g\,(\phi^\dagger \psi\chi + \psi^\dagger \phi\chi)\,,
\label{Lint}
\eeq
where $g$ is a coupling constant.  This model has been studied in great 
detail by Massar and Parentani \cite{Massar} and by Gabriel et al \cite{Gabriel},
 who used it to investigate the Unruh effect for an accelerated detector. 

We start with a charged $\psi$-particle in the 
initial state.  It interacts with $\phi$-antiparticles of the pairs 
via\footnote{Here and below we use asterisk to denote antiparticles.} $\psi\phi^* \to\chi$ and thus has a finite lifetime 
$\tau_\psi$.
We shall calculate the momentum distribution of $\chi$-particles in the final state and use it to deduce the momentum distribution of the created $\phi$-pairs.  
To achieve this goal, we will have to consider the interaction
(\ref{Lint}) with a time-dependent coupling, $g(t) = g\,\exp(-t^2/T^2)$, so that the detector is turned on for a finite period of time $\Delta t \sim T$.
We shall also briefly discuss the case when the role of the detector is played by the neutral $\chi$-particle.  All our results point in the direction of option (A) -- that the 
frame of pair (and bubble) nucleation is determined by the frame of the 
detector.

The paper is organized as follows. In Section 2, we introduce the in-vacuum state and review the Schwinger pair production using the method of Bogoliubov coefficients.
In Section 3, we show that the in-vacuum is Lorentz invariant.
In Section 4, we calculate the two-point function in this vacuum and discuss the expectation value of the current, pointing out
the pathologies associated with the infinite density of particles in the in-vacuum. We argue that these can be remedied by 
considering Lorentz breaking initial conditions. In Section 5, we set up our detector model.  The final distribution of $\chi$-particles 
and the observed momentum distribution of the pairs are calculated in Sections 6 and 7.
Finally, our results are summarized and discussed in Section 8.  {Some technical details of the calculations are presented in the Appendices.}

\section{Schwinger pair production}

We consider a constant electric field in ($1+1$)-dimensions. 
Spin-zero charged particles that are being pair produced by the Schwinger 
effect are described by a complex scalar field $\phi(t,x)$; the corresponding 
action is
\begin{eqnarray}
S&=&\int d^2x \left[
-\eta^{\mu\nu}\left(\partial_\mu+ieA_\mu\right)\phi^*
\left(\partial_\nu-ieA_\nu\right)\phi
-m^2\phi^*\phi-\frac{1}{4}F^{\mu\nu}F_{\mu\nu}
\right]\,.
\end{eqnarray}
Here, $m$ and $e$ are the mass and the electric charge of the particles,
\begin{eqnarray}
F_{\mu\nu} = \partial_\mu A_\nu-\partial_\nu A_\mu  = -\epsilon_{\mu\nu}\,E\,,
\label{constE}
\end{eqnarray}
$\epsilon_{\mu\nu}$ is the unit anti-symmetric tensor with
$\epsilon_{01}=1$, and $E={\rm const}$ is the value of the electric field.
It is clear from this expression that a constant electric field in 
($1+1$)-dimensions is Lorentz invariant. More specifically,
an observer with a 2-velocity $u^\mu$ sees the electric field given
by 
\begin{eqnarray}
E_\mu=F_{\mu\nu}\,u^\nu=E\,n_\mu\,,
\qquad n_\mu\equiv u^\nu\epsilon_{\nu\mu}\,,
\end{eqnarray}
where $n_\mu$ is a unit spacelike vector normal to $u^\mu$.

For this electric field, we may choose a Lorentz-covariant
gauge in which the gauge field $A_\mu$ is given by
\begin{eqnarray}
A_\mu=\frac{1}{2}E\,\epsilon_{\mu\nu}\,x^\nu\,.
\end{eqnarray}
However, since the calculation in this gauge seems technically
more involved, we choose a non-covariant gauge where
\begin{eqnarray}
A_\mu=\frac{1}{2}E\Bigl(\epsilon_{\mu\nu}\,x^\nu-\partial_\mu(x^0x^1)\Bigr)\,.
\end{eqnarray}
This gives the components of the gauge field as
\begin{eqnarray}
A_\mu=(A_t\,,\, A_x)=(0\,,\, -Et)\,,
\label{gauge}
\end{eqnarray}
where $(t,x)=(x^0,x^1)$.

\subsection{The in-vacuum and the Bogoliubov coefficients}

The variation of the action with respect to $\phi$ gives the field equation,
\begin{eqnarray}
\left[ 
-\partial_t^2 +\left(
\partial_x - ieA_x
\right)^2 - m^2
\right] \phi =0.
\label{eom1}
\end{eqnarray}
We expand the field in terms of the creation and anihilation operators,
\begin{eqnarray}
\phi(t,x)=\int\frac{dk}{(2\pi)^{1/2}}\left(
a_k\phi_k(t)+b_{-k}^\dag\phi_k^*(t)
\right)e^{ikx}\,,
\label{phi-field}
\end{eqnarray}
where the mode functions $\phi_k(t)$ satisfy the equation,
\begin{eqnarray}
\left[\frac{d^2}{dt^2}+m^2 +(k+eE t)^2\right] \phi_k =0\,.
\label{eom2}
\end{eqnarray}
The canonical commutation relations lead to
\begin{eqnarray}
[a_k , a_{k'}^\dag ] = \delta (k-k')  \,, \qquad 
[b_k , b_{k'}^\dag ] = \delta (k-k') \,,
\end{eqnarray}
and the normalization condition,
\begin{eqnarray}
 i\Bigl(\phi_k^* (t) \partial_t \phi_k (t) 
- \phi_k (t) \partial_t \phi_k^* (t)\Bigr)
= 1 \,. 
\label{norm}
\end{eqnarray}

Linearly independent solutions of Eq.(\ref{eom2}) can be expressed in terms of the parabolic cylinder functions,
\beq
\phi_k^{\pm}(z) \propto D_{\nu^*}[\pm (1-i)z]\,,
\eeq
where
\begin{equation}
z \equiv \sqrt{eE} \left( t + \frac{k}{eE}\right), 
\qquad 
\nu=-{1+i \lambda \over 2}  \ , \qquad \lambda\equiv \frac{m^2}{eE}.
\label{nu}
\end{equation}
The general solution is given by a linear superposition of $\phi_k^{\pm}$.  We choose
\begin{equation}
\phi_k(z)=\frac{1}{(2eE)^{1/4}}\, e^{i\frac{\pi}{4}\nu^*}D_{\nu^*} [-(1-i)  z]\,,
\label{sol:in}
\end{equation}
where the coefficient is chosen in order to satisfy the normalization condition (\ref{norm}).
For future reference, we give a useful integral representation of the parabolic cylinder functions,
\begin{eqnarray}
D_\Lambda(Z)=\frac{e^{-\frac{Z^2}{4}}}{\Gamma(-\Lambda)}
\int_0^\infty dw \,e^{-Zw-\frac{w^2}{2}}w^{-\Lambda-1}\,,\qquad
{\rm for}\qquad \Re(\Lambda)<0\,.
\label{Dintrep}
\end{eqnarray}

From the asymptotic expansion formula,
\begin{eqnarray}
D_p(z)\sim e^{-\frac{z^2}{4}} z^p\,,\qquad {\rm for}\qquad |z|\gg1\,,
\quad|z|\gg|p|\,,\quad |{\rm arg}~ z|<\frac{3}{4}\pi\,,
\end{eqnarray}
we find 
\begin{equation}
\phi_k \approx  \frac{1}{(2eE)^{1/4}} 
\left(\sqrt{2}\,|z|\right)^{\nu^* }e^{\frac{i}{2}z^2}\,, 
\qquad{\rm for}\quad
z \ll -\ |\nu| \,.
\label{inr}
\end{equation}
Hence $i\partial_t\phi_k\sim -eEt \phi_k$ at $t\to-\infty$, indicating that
these mode functions are positive frequency in the ``in" region at $t\to-\infty$.  
The corresponding vacuum state $|0\rangle_{\rm in}$, defined by
$a_k|0\rangle_{\rm in} = b_k|0\rangle_{\rm in}=0$ is the in-vacuum, which has no particles in the asymptotic ``in" region.  

Because of the non-trivial background, the positive frequency
mode function $\phi_k$ at $t\to-\infty$ does not remain positive frequency
at finite $t$, and in particular it is given by a linear combination
of the positive and negative frequency functions at $t\to +\infty$.
The positive frequency functions at $t\to\pm\infty$
are related by a Bogoliubov transformation,
\begin{eqnarray}
\phi_k = \alpha_k\,\phi_k^{\rm out} + \beta_k\,\phi_k^{{\rm out}*}\,,
\label{Bogoliubov}
\end{eqnarray}
where 
\begin{eqnarray}
\phi_k^{\rm out}(z)
= \frac{1}{(2eE)^{1/4}}\, e^{-i\frac{\pi}{4}\nu}\,D_{\nu} [(1+i)  z]\,,
\label{phiout}
\end{eqnarray}
is the positive frequency mode function
at $t\to\infty$ and $|\alpha_k|^2-|\beta_k|^2=1$.
We can check that the asymptotic expansion of $\phi_k^{\rm out}$
indeed gives
\begin{eqnarray}
\phi_k^{\rm out} \approx  
\frac{1}{(2eE)^{1/4}} \left(\sqrt{2}\,z\right)^{\nu }e^{-\frac{i}{2}z^2}, 
\qquad{\rm for}\quad
z \gg |\nu|\,. \label{outr}
\end{eqnarray}
Using a linear relation,
\begin{eqnarray}
D_{\nu^*}[-(1-i)z]=e^{i\pi\nu^*}D_{\nu^*}[(1-i)z]
+\frac{\sqrt{2\pi}}{\Gamma(-\nu^*)}\,e^{-i\frac{\pi}{2}\nu}D_\nu[(1+i)z]\,,
\end{eqnarray}
we can read off the Bogoliubov coefficients,
\begin{eqnarray}
\alpha_k = \frac{\sqrt{2\pi}}{\Gamma(-\nu^*)}\,e^{i\frac{\pi}{4} (\nu^* -\nu )}
\,,\qquad
\beta_k = e^{i\pi\nu^*}\,.
\label{bogoliubov}
\end{eqnarray}
This gives
\begin{equation}
|\beta_k|^2 = e^{-\pi \lambda} = e^{-\frac{\pi m^2}{eE}}\,.
\label{betak}
\end{equation}

\subsection{Momentum distribution and Lorentz invariance}

The Bogoliubov coefficients $\beta_k$ are simply related to the momentum distribution of particles in the ``out" state at $t\to\infty$,
\begin{equation}
\frac{dn}{dk} = \frac{1}{2\pi} |\beta_k|^2\,.
\label{dndk}
\end{equation}
This can in turn be related to the rate of the Schwinger process,
ie, the rate of pair creation of charged particles.
The mixing between positive and negative frequency 
modes in the in-vacuum mode functions (\ref{sol:in}) occurs in the interval 
\begin{equation}
|z|\lesssim |\nu|\,, \label{zmixing}
\end{equation} 
centered at $z=0$ (or $k=-eE t$). 
Hence, the number density of particles created between time $t_0$ and $t>t_0$ 
is given by
\begin{equation}
n = \frac{1}{2\pi}\int^{-eEt_0}_{-eEt} dk\  |\beta_k|^2\,.
\end{equation}
This leads to
\begin{equation}
\frac{dn}{dt} = \left(\frac{eE}{2\pi}\right) e^{-\frac{\pi m^2}{eE}}\,.
\label{dndt}
\end{equation}
This is the Schwinger formula for the rate of pair creation.

An important feature of the momentum distribution (\ref{dndk}) is that it is independent of $k$.
The total density of created particles, obtained by integration over $k$, is infinite.  This is not surprising, since the pair creation process was going on at a constant rate for an infinite time.
A more realistic calculation would include back-reaction of the pairs on the electric field, so the field would gradually decrease and only a finite density of pairs would be produced.  But according to Eq.~(\ref{dndt}), in a weak electric field with $|eE|\ll m^2$ pair creation is a very slow process, and the field can remain nearly constant for a very long time.  Thus, one can expect that our idealized treatment should apply in some limiting sense.  We shall see, however, that taking the limit and interpreting the result is not always straightforward.   

It should be noted that $k$ in Eq.~(\ref{dndk}) is the canonical momentum, which is related to the physical momentum by
\beq
k_{\rm phys}=k+eEt.
\label{kphys}
\eeq
The only subtlety is in determining the range of this distribution. Pair creation in a given mode occurs in a 
spacetime region where the negative frequency contribution to the mode function becomes significant.  As we discussed, this occurs at $k\approx -eE t$. Hence,
Eq. (\ref{dndk}) is valid in the range 
\begin{equation}
-eE t \lesssim k < \infty. \label{cut}
\end{equation}
The uncertainty in the lower limit of this range can be estimated from (\ref{zmixing}), 
which (assuming $\lambda=m^2/(eE)  \gg 1$) leads to the uncertainty 
\begin{equation}
\Delta k_{\rm min} \sim \lambda^{1/2} m\,,
\qquad
{\rm for}\quad \lambda \gg 1\,.
\label{kmin}
\end{equation} 
Note that at future infinity, the distribution of out particles is given by (\ref{dndk}) in the full range $-\infty <k<\infty$. 
The final distribution is then Lorentz invariant\footnote{This conclusions follows from the Lorentz invariance of the phase space element $dk\,dx$
\cite{Landau}.}.
On the other hand, at any finite time, the invariance is broken by the lower cut-off in Eq. (\ref{cut}).

Eq.~(\ref{kphys}) leads to $dk_{\rm phys} = dk$, and since $\beta_k$ are independent of $k$, 
we have
\begin{equation}
\frac{dn}{dk_{\rm phys}} = \frac{1}{2\pi} |\beta_k|^2\,.
\label{dndkphys}
\end{equation}
From (\ref{kphys}) and (\ref{cut}), the range of physical momentum for the created particles is  
\begin{equation}
0 \lesssim  k_{\rm phys} < \infty.   \label{rangekphys}
\end{equation}
Hence, we expect to see particles which predominantly have positive momentum
\begin{equation}
\frac{dn}{dk_{\rm phys}} \approx \frac{1}{2\pi} |\beta_k|^2\,
\theta\left(k_{\rm phys}\,\right) \,,
\label{theta}
\end{equation}
and similarly
\begin{equation}
\frac{dn_*}{dk_{\rm phys}} \approx \frac{1}{2\pi} |\beta_k|^2\,
\theta\left(-k_{\rm phys}\,\right) \,
\label{thetaanti}
\end{equation}
for antiparticles.
This restriction on $k_{\rm phys}$ is intuitively plausible.  
Pairs are produced at rest; then particles and antiparticles are accelerated by the field in opposite directions.  
In the limit of infinite time, one can expect to find particles having arbitrarily large momentum, but one would not find particles with $k_{\rm phys}<0$ (or antiparticles with $k_{\rm phys}>0$)\footnote{Note, however, that the uncertainty in the lower limit for $k_{\rm phys}$ is of order $\Delta k_{\rm phys,min} \sim \lambda^{1/2} m$. 
For $\lambda \gg 1$, this seems to leave some room for highly relativistic particles with both signs of the physical momentum.\label{footlambda}}.

A potential problem with this picture is that the cutoff at $k_{\rm phys}\approx 0$ clearly breaks Lorentz invariance.  This is not necessarily impossible. On one hand, even though the background is Lorentz invariant, 
the invariance could be broken by the in-vacuum state which we are using for the $\phi$-field. This would correspond to option (C) in the Introduction, where the frame of nucleation might be
influenced by  how the decaying vacuum is set up.  Finally, we must consider the possibility that the in-vacuum is Lorentz invariant, but all observers see the distributions 
(\ref{theta}) and (\ref{thetaanti}) for the momenta
of particles and anti-particles relative to the detector's rest frame. This would correspond to option (A) in the Introduction.




 Our next task is therefore to investigate the Lorentz invariance of the adiabatic in-vacuum.
 
\section{Lorentz invariance of the in-vacuum}





To examine the behavior of the in-vacuum under Lorentz transformations, we
compare two in-vacua defined with respect to different rest
frames which are related by a Lorentz boost.
To do so, we consider a gauge transformation that relates the
components of the gauge field in the two frames, such that
$A_\mu$ has the same form given by Eq.~(\ref{gauge}) in each
frame.
A positive frequency mode function of
$\phi$ at $t\to-\infty$ in the gauge (\ref{gauge}) reads
\begin{eqnarray}
\phi_k(z)\,e^{ikx}\,;\qquad
z=\sqrt{eE}\left(t+\frac{k}{eE}\right)\,,
\label{pfreq}
\end{eqnarray}
with $\phi_k(z)$ from Eq.~(\ref{sol:in}).  We consider another Lorentz frame given by
\begin{eqnarray}
\bar t=\gamma(t-vx)\,,\qquad
\bar x=\gamma(x-vt)\,;
\qquad \gamma=(1-v^2)^{-1/2}\,.
\label{Ltrans}
\end{eqnarray}
In the barred frame, if we adopt the gauge $\bar A_{\bar0}=0$, 
the positive frequency modes at ${\bar t}\to-\infty$ are
\begin{eqnarray}
{}_0\bar\phi_{\bar k}(\bar z)\,e^{i\bar k\bar x}\,;
\qquad\bar z=\sqrt{eE}\left(\bar t+\frac{\bar k}{eE}\right)\,,
\end{eqnarray}
where the prefix `0' means the gauge 
\begin{eqnarray}
\bar A_{\bar0}=0\,.
\label{bA0gauge}
\end{eqnarray}
Note that the functions ${}_0\bar\phi_{\bar k}(\bar z)$
are the same as $\phi_k(z)$ with the replacement $z\to\bar z$,
\begin{eqnarray}
{}_0\bar\phi_{\bar k}(\bar z)=\phi_{\bar k}(\bar z)\,.
\end{eqnarray}

To compare the positive frequency function in the barred frame
with the one in the original unbarred frame,
we have to express the barred frame positive frequency fuction
in terms of the original coordinates $(t,x)$. To do so, we
perform a gauge trasformation from $\bar A_{\bar0}=0$ to $A_{0}=0$ gauge.

The electromagnetic vector potential trasforms from the barred
frame to the original frame as
\begin{eqnarray}
{}_0A_{\mu}=\frac{\partial \bar x^\nu}{\partial x^\mu}\bar A_{\bar\nu}\,,
\end{eqnarray}
where ${}_0A_\mu$ is the components of the potential
in the gauge (\ref{bA0gauge}) in the coordinates $(t,x)$.
This gives
\begin{eqnarray}
&&{}_0A_{0}=\frac{\partial \bar x}{\partial t}\bar A_{\bar1}
=\gamma v E\,\bar t=\gamma^2 v E(t-vx)\,,
\cr
&&{}_0A_1=\frac{\partial \bar x}{\partial x}\bar A_{\bar1}
=-\gamma E\,\bar t=-\gamma^2 E(t-vx)\,.
\end{eqnarray}
We perform the gauge transformation from this gauge to the 
gauge (\ref{gauge}),
\begin{eqnarray}
A_\mu={}_0A_\mu-\partial_\mu\Omega\,,
\end{eqnarray}
so that $A_\mu$ is given by (\ref{gauge}). Namely we set
\begin{eqnarray}
A_0={}_0A_0-\partial_0\Omega=0\,, \quad
 A_1={}_0A_1-\partial_1\Omega=-E\,t\,.
\end{eqnarray}
This can be easily solved. We find
\begin{eqnarray}
\Omega=\frac{E\gamma^2v}{2}\left(t^2-2vtx+x^2\right)\,.
\label{Omega}
\end{eqnarray}
Under this gauge transformation, since the covariant
derivative of $\phi$,
\begin{eqnarray}
\left(\partial_\mu-ieA_\mu\right)\phi\,,
\end{eqnarray}
 is gauge-invariant, the mode function trasforms as
\begin{eqnarray}
{}_0\bar \phi_{\bar k}\to \bar \phi_{\bar k}
={}_0\bar \phi_{\bar k}\,e^{-ie\Omega}\,.
\label{gtU}
\end{eqnarray}
Thus the positive frequency mode function in the barred
frame is expressed in terms of the original coordinates as
\begin{eqnarray}
\bar\phi_{\bar k}\,e^{i\bar k\bar x}\,;
\qquad
\bar\phi_{\bar k}(t,x)=\phi_{\bar k}(\bar z)\,e^{-ie\Omega(t,x)}\,,
\end{eqnarray}
where
\begin{eqnarray}
\bar z=\sqrt{eE}\gamma(t-vx)+\frac{\bar k}{\sqrt{eE}}\,,
\qquad
\bar x=\gamma(x-vt)\,.
\label{barz}
\end{eqnarray}


A simple way to examine the Lorentz invariance of the in-vacuum
is to evaluate the Klein-Gordon inner product 
\begin{eqnarray}
K\equiv\int_{t={\rm const}}\!\! dx\,
\left(\,\phi_k\,\partial_t\bar \phi_{\bar k}-\bar \phi_{\bar k}\,\partial_t \phi_k\,\right) e^{-ikx+i\bar k \bar x}
\label{KGdef}
\end{eqnarray}
This product characterizes the mixing of positive with negative frequency modes. 
If $K$ is non-vanishing, it means that the vacua defined by
$\bar \phi_{\bar k}$ and by $\phi_k$ are intrinsically different,
hence the in-vacuum is not Lorentz invariant. 
The inner product (\ref{KGdef}) is computed in Appendix~\ref{apx:KGproduct} 
using the exact mode functions, and the result is
\begin{equation}
K=0\,.
\end{equation}
This shows that the in-vacuum (hence also the out-vacuum)
is Lorentz invariant. 

We shall see in the next Section that this state has certain unphysical features. These would actually make it
pathological when we consider interactions beyond the tree level.

\section{Two-point function, Current, and Lorentz breaking quantum states}

\label{sectioncurrent}

A quantity of particular interest for our problem is the expectation value of the electric current $\langle J_\mu \rangle$.
The physical picture of particles and antiparticles produced at a constant rate and driven in opposite directions by the electric field, suggests that there should be a steadily growing electric current,
\beq
\frac{\partial \langle J_1\rangle}{\partial t} = C\,,
\label{dJ1dt}
\eeq
where $C={\rm const}$, while the charge density remains equal to zero,
\beq
\langle J_0 \rangle =0\,.
\label{J0}
\eeq
On the other hand, the expectation value of any vector in a Lorentz invariant quantum state should be zero.  Since we found that the in-vacuum is Lorentz invariant, this indicates that we should have 
$\langle J_\mu \rangle =0$.  Once again, there is some tension between Lorentz invariance and the physical picture of pair production. Let us investigate this question in some detail.


Formally, the current can be expressed as
\beq
J_\mu = \frac{ie}{2} \left(\phi^\dagger D_\mu\phi - \phi \,(D_\mu \phi)^\dagger \right) + {\rm h.c.} 
\label{currentform}
\eeq
A naive calculation of the expectation value of $J_\mu$ in the in-vacuum gives an infinite answer.  
One way to regulate this sort of infinity, which respects Lorentz and gauge invariance, is to use the point splitting method.  The idea is to take the two field operators in the products in Eq.~(\ref{currentform}) at different spacetime points, $x$ and $y$, and then take the limit $y\to x$.
The current expectation value $\langle J_\mu \rangle$ can then be expressed in terms of the 2-point function,
\begin{eqnarray}
G^{(1)}=G^+ + G^-\,, 
\end{eqnarray}
where
\begin{equation}
G^+(x^\mu,y^\mu) \equiv {}_{\rm in}\langle\,0\,|\,
\phi^\dagger(x^\mu)~ e^{-ie \int_{x}^{y} A_\nu dx^\nu} ~\phi(y^\mu)\,
|\,0\,\rangle_{\rm in}\,,
\label{G+}
\end{equation}
\begin{equation}
G^-(x^\mu,y^\mu)\equiv {}_{\rm in}\langle\,0\,|\,
\phi(y^\mu)~ e^{-ie \int_{x}^{y} A_\nu dx^\nu}~ \phi^\dagger (x^\mu)\,
|\,0\,\rangle_{\rm in}\,,
\label{G-}
\end{equation}
and the Wilson line has been inserted to make the 2-point function 
gauge invariant.  The expectation value of 
the current is given by
\begin{equation}
\langle J_\mu \rangle = {i e\over 2} 
\lim_{x^\nu \to y^\nu} 
\left({\partial\over \partial y^{\mu}} 
- {\partial\over \partial x^{\mu}}\right) 
G^{(1)}(y^\nu-x^\nu) = 
i e \lim_{\Delta x^\nu \to 0}
{\partial G^{(1)}(\Delta x^\nu) \over \partial \Delta x^\mu}\,,
\label{JG}
\end{equation}
where $\Delta x^\nu\equiv y^\nu-x^\nu$.

The 2-point functions $G^{\pm}(\Delta x^\mu)$ are calculated in Appendix~\ref{apx:2pfunction}.
The result is
\begin{equation}
G^+(\Delta x^\mu)=\frac{|\alpha|^2}{4\pi}
 \Gamma(-\nu^*)\,
 \frac{W_{i\lambda/2,\,0}\left(i eE\,\Delta s^2/2\right)}
{\sqrt{i eE\,\Delta s^2/2}}\,, 
\qquad {\rm for} \qquad
\Delta t-\Delta x >0 
\label{60}
\end{equation}
and 
\begin{equation}
G^+(\Delta x^\mu)=\frac{|\alpha|^2}{4\pi}
 \Gamma(-\nu)\,
\frac{W_{-i\lambda/2,\,0}\left(-i eE\,\Delta s^2/2\right)}
{\sqrt{-i eE\,\Delta s^2/2}}\,,
\qquad {\rm for} \qquad
\Delta t-\Delta x < 0 . 
\label{61}
\end{equation}
Here, $W_{\sigma,\,\rho}\,(z)$ are Whittaker functions, $\Delta s^2= -(\Delta t)^2+(\Delta x)^2$, and $\alpha$ is the Bogoliubov coefficient given by Eq.~(\ref{bogoliubov}).  We have omitted the index of momentum $k$, since $\alpha$ is independent of $k$.  The function $G^{-}$ is simply related to $G^{+}$,
\begin{equation}
G^-(\Delta t,\Delta x) = G^+(-\Delta t,\Delta x).
\end{equation}
For $\Delta t + \Delta x < 0$ this is given by (\ref{60}),
 and for $\Delta t + \Delta x > 0$ it is given by (\ref{61}).

The first thing we note is that the 2-point functions $G^\pm$ and $G^{(1)}$ are Lorentz invariant.  This is a further manifestation of the Lorentz invariance of the in-vacuum.  The function $G^{(1)}$ is also invariant with respect to time reversal,
\begin{equation}
G^{(1)}(\Delta t,\Delta x) = G^{(1)}(-\Delta t,\Delta x),
\end{equation}
but not with respect to spatial reflection,
\begin{equation}
G^{(1)}(\Delta t,-\Delta x) = G^{(1)}(\Delta t,\Delta x)^*.
\end{equation}
The latter is not surprising, since the electric field defines a preferred direction on the $x$-axis.  $G^{(1)}$ is invariant under a simultaneous transformation $\Delta x\to -\Delta x$, $E\to -E$.
Note also that the functions $G^\pm$ are not invariant under time reversal; hence the in-vacuum state is not time reversal invariant, as one might expect.

For the calculation of the current (\ref{JG}), we are interested in the 2-point functions in the limit of $\Delta x^\mu \to 0$.  This is given by
\beq
G^+(\Delta x^\mu)\approx -\frac{|\alpha|^2}{4\pi}\left[ \ln\left(\frac{ieE\Delta s^2}{2}\right) +\psi \left(\frac{1}{2} - \frac{i\lambda}{2}\right)+2\gamma\right] 
\label{G+asymp}
\eeq
for  $\Delta t - \Delta x > 0$ and by a complex conjugate expression for  $\Delta t - \Delta x < 0$.
Here, $\gamma$ is Euler's constant.  The leading term in the limit of large mass, $m^2\gg |eE|$, is
\beq
G^\pm (\Delta x^\mu)\approx -\frac{|\alpha|^2}{4\pi} \ln\left(\frac{m^2\Delta s^2}{4}\right).
\eeq

The singularity structure of the 2-point functions $G^\pm$, and therefore of $G^{(1)}$ is rather unusual.  Most importantly, the leading divergent term at $\Delta s^2 \to 0$  is different from that for a free field: the coefficient multiplying the logarithm in (\ref{G+asymp}) has an extra factor of $|\alpha|^2$, compared to the case of $E=0$.  In terms of the standard terminology, the 2-point function is not of the Hadamard form.  Another non-Hadamard feature is the $\theta$-function discontinuity arising from the different forms of $G^\pm$ at $\Delta t = \pm \Delta x$.  

The non-standard singularity structure of the 2-point functions makes the regularization of the expectation values of the current $J_\mu$ and of the energy-momentum tensor $T_{\mu\nu}$ problematic.  The point splitting method with Hadamard subtraction cannot be used, and even if one does not insist on the Hadamard form of the 2-point functions, the coincidence limits of $G^{(1)}$ and of its derivatives are ill defined, because of the discontinuity on the light cone.  The Pauli-Villars regularization also appears to fail, since the coefficient multiplying the leading divergence in (\ref{G+asymp}) is mass-dependent.  This seems to indicate
that Lorentz invariant regularization of the expectation values of observables is impossible in our in-vacuum.

As mentioned in the introduction, this is to be expected, since the Lorentz invariant in-state contains an infinite number of created pairs. In other words, the divergent current is caused by ``actual" particles, corresponding to
the non-vanishing occupation numbers of all momentum states with $k+eEt > 0$ 
(rather than by zero point fluctuations, whose effect can be subtracted by using Lorentz invariant counterterms).

In Appendix~\ref{apx:LVcurrent} we calculate the expectation value of $J_\mu$ using
a direct momentum space regularization,
\begin{equation}
|k| \ll k_{\rm max}\,, \label{kmax}
\end{equation}
and then taking the limit $k_{\rm max} \to \infty$. Physically, the current is due to the created pairs, and therefore the regulator (\ref{kmax}) 
amounts to ignoring the contribution from pairs with $|k|>k_{\rm max}$.  Noting that the mixing between positive and negative frequency modes occurs near the time 
when $k_{\rm phys} = k + eE t = 0$, we can think of (\ref{kmax}) as a proxy for the situation where the electric
field is only turned on for a finite interval of time\footnote{The expectation value of the current with an electric field turned on for a finite period of time has been studied by a number of authors (e.g., Refs.\cite{Novikov:1980ni, Krotov:2010ma, Gavrilov:2007hq, Gavrilov:2012jk}).  In this case, Lorentz invariance is explicitely broken and the total number of produced pairs is finite.  The rate of current growth during the period when the field is nearly constant agrees with our result (\ref{dJ1dt}), (\ref{Cfrac}).}, $|t| < k_{\rm max}/(eE)$.
The regulator (\ref{kmax}) is not manifestly Lorentz invariant and, not surprisingly, the result which is obtained by using it is not Lorentz invariant either; 
it agrees with Eqs.~(\ref{dJ1dt}),(\ref{J0}) with
\beq
C = \frac{e^2 E}{\pi} e^{-\frac{\pi m^2}{eE}}\,.
\label{Cfrac}
\eeq
This is what one would expect on physical grounds: the rate of current growth is proportional to the rate of Schwinger pair production.


In view of the unphysical properties of the Lorentz invariant in-vacuum, 
we are led to the conclusion that (in the presence of a constant electric field), no physical state can be Lorentz invariant.

To illustrate this point more explicitly, we may consider the closely related example 
of massless spinor QED in $1+1$ dimensions \cite{Peskin}. 
Aside from the conserved electric current
$J^\mu = e{\bar\psi}\gamma^\mu \psi$, this model also has an axial current,
which in $1+1$ dimensions is related to the electric current by
\beq 
{\tilde J}^\mu = e{\bar\psi}\gamma^\mu \gamma^5 \psi = \epsilon^{\mu\nu}J_\nu\,. \label{currents}
 \eeq
Here, $\gamma^5 \equiv\gamma^0 \gamma^1$.
As is well known  the axial current would be classically conserved, but
any gauge invariant (and Lorentz invariant) regulator will lead to a nonzero divergence
given by the anomaly equation \cite{Adler,Jackiw}:
\beq
\partial_\mu {\tilde J}^\mu  = \frac{e^2 E}{\pi}\,.
\label{anomaly}
\eeq
This equation is of course valid even if the electric field is not constant, but here we are primarily interested in the
case where it is, and hence the background is Lorentz invariant. Combining Eqs. (\ref{currents}) and (\ref{anomaly}),
the renormalized electric current satisfies
\begin{equation}
 \epsilon^{\mu\nu}\partial_\mu J_\nu = \frac{e^2 E}{\pi}\,. \label{spontaneous}
 \end{equation}
This is consistent with our Eqs.~(\ref{dJ1dt}),(\ref{J0}),(\ref{Cfrac}) in the massless limit $m=0$.
 Note that even though Eq.~(\ref{spontaneous}) for the current has a covariant form, it does not admit Lorentz-invariant 
 solutions.  In this sense, Lorentz invariance is spontaneously broken:
quantum states satisfying Eq.~(\ref{spontaneous}) cannot be Lorentz invariant.

\section{A model detector}

We now turn to the question of how the pair production process is seen by different observers.  To this end, we introduce a model detector which involves two additional scalar fields: a charged scalar $\psi(t,x)$ and a neutral scalar $\chi(t,x)$, with the interaction
\begin{eqnarray}
S_{\rm{int}}=-g\int d^2x \left[
\phi^\dag\psi\chi
+\psi^\dag\phi\chi
\right]
= - \int dt H_{\rm int}\,,
\label{interaction}
\end{eqnarray}
where $g$ is a coupling constant.  We denote the masses of $\phi$,$\psi$ and $\chi$ particles by
$m_\phi$, $m_\psi$ and $m_\chi$, respectively.  The charge of $\psi$-particles is the same as that of $\phi$-particles, as required by gauge invariance of the action. 

Suppose first that we have a charged $\psi$-particle in the initial state.  It will scatter on a $\phi$-antiparticle of the pairs, producing a $\chi$ in the final state.  Such an interaction represents a ``detection" of $\phi^*$, with $\phi^*$ playing the role of a domain wall in our model.
Alternatively, we can have a neutral $\chi$ particle in the initial state, in which case the detection process yields a $\psi$-particle in the final state.

It will be useful to consider the kinematics of the scattering process.  Starting with
\beq
\psi\phi^*\to\chi,
\label{psiphichi}
\eeq  
let $q$, $-k$ and $p$ be the momenta of $\psi$, $\phi^*$ and $\chi$, respectively. Momentum conservation requires that
\beq
q-k=p.
\label{Pcons}
\eeq
We remind the reader that $k$ is the canonical momentum of $\phi$, which is related to the physical momentum by $k_{\rm phys} = k+eEt$ (and opposite sign for $\phi$-antiparticle).  Similarly, for $\psi$-particle, $q_{\rm phys} = q+eEt$.  For the neutral $\chi$-particle, $p_{\rm phys}=p$.  Hence, Eq.~(\ref{Pcons}) implies
\beq
q_{\rm phys}-k_{\rm phys}=p_{\rm phys},
\label{Pphyscons}
\eeq 
which simply states the momentum conservation at the moment of collision.

In the absence of an electric field, we also have energy conservation,
\beq
\sqrt{q^2_{\rm phys} + m_\psi^2}+ \sqrt{k^2_{\rm phys} + m_\phi^2} = \sqrt{p^2 + m_\chi^2}.
\label{Econs}
\eeq
Squaring this equation and combining with (\ref{Pcons}), we obtain, after some algebra,
\beq
2m_\psi^2\ k_{\rm phys} =M^2\ q_{\rm phys} \pm \sqrt{q^2_{\rm phys}+m_\psi^2}\,\sqrt{M^4 - 4m_\phi^2 m_\psi^2},
\label{Mmm}
\eeq
where
\beq
M^2\equiv m_\chi^2-m_\psi^2-m_\phi^2.
\eeq
For a given momentum of the probe $q_{\rm phys}$, there are two values of $k_{\rm phys}$ for which scattering is possible: one corresponding to the $\phi$-particle coming from the left and the other from the right.  
Note also that the scattering is kinematically allowed only if
\beq
m_\chi > m_\psi + m_\phi.
\eeq
Similar relations can be written for the neutral probe scattering $\chi\phi\to\psi$.  The corresponding relation for the masses is $m_\psi > m_\chi + m_\phi$.

In the presence of an electric field, the energy conservation (\ref{Econs}) does not strictly apply.  One way to think about this is that the interaction process is not momentary and involves some time uncertainty $\delta t$.  Charged particles are accelerated by the electric field during this time interval, so their energy is changed.  Alternatively, energy non-conservation can be understood in terms of the Unruh effect \cite{Massar,Gabriel}.  The charged detector particle $\psi$ moves with an acceleration $a=eE/m_\psi$ and is exposed to an effective temperature $T=a/(2\pi)$.  Processes with energy  fluctuation $\Delta E$ can therefore occur with a Boltzmann suppressed probability $\propto \exp(\Delta E/T)$.  For example, a $\psi$-particle can decay, $\psi\to\phi\chi$, even when $m_\psi < m_\phi + m_\chi$, so the decay would be kinematically forbidden in the absence of an electric field.



Apart from the $\phi$-pair production, $\psi$-particle pairs will also be produced by the electric field.  In addition, triplets of $\phi\psi^*\chi$ and of $\phi^*\psi\chi$ will be spontaneously produced from the vacuum.  Various interaction processes in our model are illustrated in Fig.~1.  
In order to avoid false detections, we have to choose the parameters of the model so that processes shown in Figs.~1(a) and 1(c) are suppressed compared to the true detection process shown in Fig.~1(b).  This can be achieved by requiring that 
\begin{eqnarray}
m_\psi \gg m_\phi \,;
\qquad (m_\chi - m_\psi)m_\psi \gg m_\phi^2\,.
\label{suppression}
\end{eqnarray}
The first and the second of these conditions ensure the suppression of processes 1(a) and 1(c), respectively.

\begin{figure}
\begin{center}
\vspace{-2cm}
\includegraphics[width=14cm]{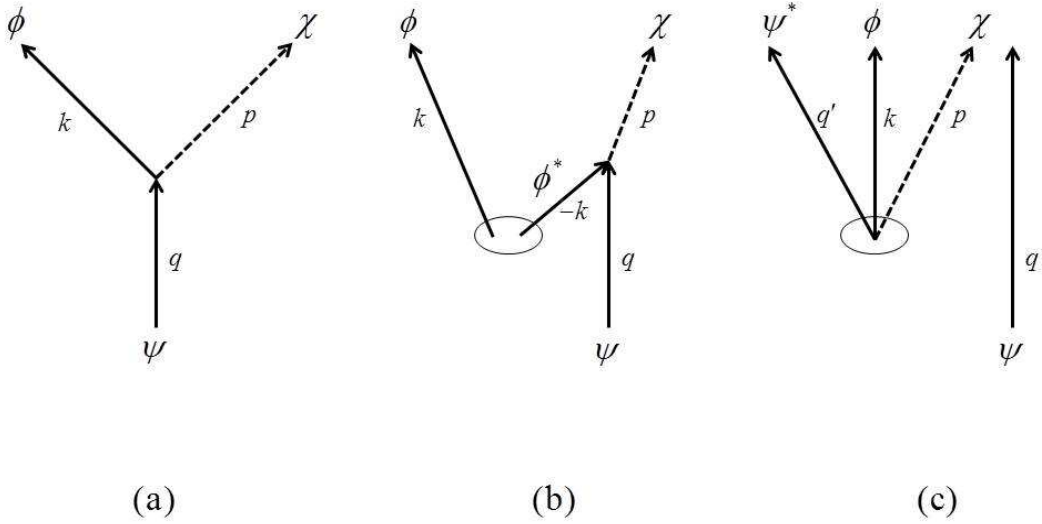}
\vspace{-1.5cm}
\caption{Interaction processes in the $\phi\psi\chi$ model with a $\psi$-particle in the initial state.  
(a) Decay of $\psi$ into $\phi$ and $\chi$.  In the presence of an electric field, this can occur even if $m_\psi < m_\phi + m_\chi$.  (b) True detection process: $\psi$ collides with a $\phi^*$ coming from a $\phi\phi^*$ pair, producing a $\chi$.  (c) A triplet of $\psi^*$, $\phi$ and $\chi$-particles is spontaneously produced from the vacuum.  The initial $\psi$-particle does not participate in the interaction.}
\end{center}
\end{figure}

We shall calculate the amplitudes for the detection processes like (\ref{psiphichi}) to lowest order of perturbation theory in the coupling $g$, but exactly with respect to the background electric field $E$. 
We first expand all the fields in terms of the in-vacuum creation
and annihilation operators as we did in (\ref{phi-field}),
\begin{eqnarray}
\phi&=&
\int\frac{dk}{(2\pi)^{1/2}}\left(a_k\phi_k+b_{-k}^\dag\phi_k^*\right)e^{ikx}\,,
\qquad
\phi^\dag=
\int\frac{dk}{(2\pi)^{1/2}}\left(b_{-k}\phi_k+a_k^\dag\phi_k^*\right)e^{-ikx}\,,
\cr
\psi&=&\int\frac{dq}{(2\pi)^{1/2}}\left(d_q\psi_q+f_{-q}^\dag\psi_q^*\right)e^{iqx}\,,
\qquad
\psi^\dag
=\int\frac{dq}{(2\pi)^{1/2}}\left(f_{-q}\psi_q+d_q^\dag\psi_q^*\right)e^{-iqx}\,,
\cr
\chi&=&\int\frac{dp}{(2\pi)^{1/2}}\left(c_p\chi_p+c_{-p}^\dag\chi_p^*\right)e^{ipx}\,.
\label{fields}
\end{eqnarray}
Here, $\phi_k$, $\psi_q$ and $\chi_p$ are positive frequencty mode functions
of the respective fields at $t=-\infty$.  {For $\phi$ and $\psi$ fields they are given by Eq.~(\ref{sol:in}) with masses $m_\phi$ and $m_\psi$, respectively.  For the $\chi$ field,
\beq
\chi_p = \frac{1}{\sqrt{2\omega_p}}\,e^{ipx - i\omega_p t},
\eeq
where $\omega_p = \sqrt{p^2 + m_\chi^2}$.}  Since $\chi$ is neutral, $\chi_p$ is also positive frequency at $t=+\infty$.  The canonical commutation relations are
\begin{eqnarray}
&&[a_k , a_{k'}^\dag ] = \delta (k-k')  \,, 
\qquad 
[b_k , b_{k'}^\dag ] = \delta (k-k') \,,
\nonumber\\[0.5mm]
&&[d_q , d_{q'}^\dag ] = \delta (q-q')  \,, 
\qquad~ 
[f_q , f_{q'}^\dag ] = \delta (q-q')\,,
\nonumber\\[0.5mm]
&& 
[c_p , c_{p'}^\dag ] = \delta (p-p')\,.
\label{commutation}
\end{eqnarray}

In the interaction picture, the initial state $|\,{\rm i}\,\rangle$ at $t\to -\infty$ evolves to a final state $|\,{\rm f}\,\rangle$ at $t\to +\infty$, which is given by
\begin{eqnarray}
|\,{\rm f}\,\rangle &=&\exp [ -i\int_{-\infty}^\infty dt' H_{\rm int}(t') ]
\ |\,{\rm i}\,\rangle
\cr
&\simeq&
|\,{\rm i}\,\rangle - i\int_{-\infty}^\infty dt' 
H_{\rm int}(t')\,|\,{\rm i}\,\rangle \,,
\end{eqnarray}
where in the second line we kept only terms up to linear order in $g$.

With a $\psi$-particle detector, the initial state is
\beq
|\,q\,\rangle=d_q^\dag~ |\,0\,\rangle\,,
\label{initialstate}
\eeq
where $|\,0\,\rangle$ stands for the in-vacuum state, and we can examine the detection process by studying  
the distribution of $\chi$-particles in the final state, 
\begin{eqnarray}
\frac{dN_\chi}{dp}=
\frac{\langle\,{\rm f}\,|~ c_p^\dag \,c_p ~|\,{\rm f}\,\rangle}{\langle\, q\,|\,q\,\rangle}
\label{expctnum}
\end{eqnarray}
To the lowest order in $g$, the out-expectation value in (\ref{expctnum}) can be expressed as
\begin{eqnarray}
\langle\,{\rm f}\,|~ c_p^\dag\, c_p ~|\,{\rm f}\,\rangle
&=&
\int_{-\infty}^\infty dt' \int_{-\infty}^\infty dt'' 
\int
dk_1\,
dp_1
\int
dk'_1\,
dp'_1
\cr
&\times&
\langle\,q\,| H_{\rm int}(t'') | p'_1, k'_1 \rangle~
\langle p'_1, k'_1 |~ c_p^\dag\, c_p ~| p_1, k_1 \rangle~
\langle p_1, k_1 | H_{\rm int}(t') |\,q\,\rangle ,
\label{expctnum1}
\end{eqnarray}
where we have inserted two sets of intermediate states defined in the in-vacuum,  
\beq
|p',k'\rangle = c_{p'}^\dagger\,a_{k'}^\dagger |\,0\,\rangle.
\eeq
Strictly speaking, we also had to include intermediate states of the form
\beq
d_q^\dagger\,f_{q'}^\dagger\,c_{p'}^\dagger\,a_{k'}^\dagger\,|\,0\,\rangle
\eeq
and
\beq
d_q^\dagger\,d_{q'}^\dagger\,c_{p'}^\dagger\,b_{k'}^\dagger\,|\,0\,\rangle\, ,
\eeq
which would also give non-vanishing matrix elements of $H_{\rm int}$.  Inclusion of these states would account for the $\phi\psi^*\chi$ and $\phi^*\psi\chi$ triplet creation processes of Fig.~1(c).  We disregard these processes, assuming that the conditions (\ref{suppression}) are satisfied.

Using Eqs. (\ref{commutation}), we have
\begin{eqnarray}
\langle p'_1, k'_1 |~ c_p^\dag\, c_p ~| p_1, k_1 \rangle
= \delta(p-p'_1)\,\delta(p_1-p)\,\delta(k_1-k'_1) \ ,
\end{eqnarray}
and Eq.~(\ref{expctnum1}) simplifies to
\begin{eqnarray}
\langle\,{\rm f}\,|~ c_p^\dag\, c_p ~|\,{\rm f}\,\rangle
&=&
\int
dk~
\left|{\cal M}(p,q,k)\right|^2\,,
\label{expctnum2}
\end{eqnarray}
where
\begin{eqnarray}
{\cal M}(p,q,k)=\int_{-\infty}^\infty dt'\,
\langle p, k | H_{\rm int}(t') |\,q \,\rangle  \ .
\label{A}
\end{eqnarray}

Substituting the mode expansions (\ref{fields}) into Eq.~(\ref{interaction}) for $H_{\rm int}$, we obtain
\beq
{\cal M}(p,q,k) = \frac{g}{\sqrt{2\pi}}~\delta(q-p-k)\int_{-\infty}^{\infty}dt'~\phi_{k}^*\,\psi_q \,\chi_p^* \ .
\label{A1}
\eeq
Here, the delta-function enforces the momentum conservation, $p=q-k$.
Substituting this in Eq.~(\ref{expctnum2}), we find
\beq
\langle\,{\rm f}\,|~ c_p^\dag\, c_p ~|\,{\rm f}\,\rangle =
\frac{g^2}{2\pi}~\delta(0)\left|~
\int_{-\infty}^{\infty}dt~\phi_{q-p}^*~\psi_q~\chi_p^*
~\right|^2\,.
\eeq
Finally, substituting into (\ref{expctnum}) and taking into account the normalization
\begin{eqnarray}
\langle\, q\,|\,q\,\rangle = \delta(0)  \ ,
\end{eqnarray}
we obtain the momentum distribution of $\chi$-particles in the out-state,
\begin{eqnarray}
\frac{dN_\chi}{dp}=
\frac{1}{2\pi}\left|~
g \int_{-\infty}^{\infty}dt~\phi_{q-p}^*~\psi_q~\chi_p^*
~\right|^2  
\equiv \frac{1}{2\pi}  \left|{\cal A}_\chi (p;q)\right|^2 \ .
\label{expctnum3}
\end{eqnarray}

In the alternative setup, with a neutral $\chi$-particle in the initial state, 
the final distribution of $\psi$-particles is given by
\beq
\frac{dN_\psi}{dq} = \frac{\langle\,{\rm f}\,|~ {d_q^{\rm out}}^\dagger\,d_q^{\rm out} ~|\,{\rm f}\, \rangle} {\langle\,p\,|\,p\,\rangle} ,
\eeq
where $|\,p \,\rangle = c_p^\dagger \,|\,0\,\rangle$ and ${d_q^{\rm out}}^\dagger$, $d_q^{\rm out}$ are the creation and annihilation operators for $\psi$-particles in the out-vacuum, which are related to the in-vacuum operators by Bogoliubov transformations.  Assuming that $m_\psi \gg m_\phi$, $\psi$-pair production can be neglected, so we can disregard the distinction between the in- and out-vacua for $\psi$.  Then, following the same steps as above, we obtain
\begin{eqnarray}
\frac{dN_\psi}{dq}=
\frac{1}{2\pi}\left|~
g \int_{-\infty}^{\infty}dt~\phi_{q-p}^*~\psi_q^*~\chi_p
~\right|^2  
\equiv \frac{1}{2\pi}  \left|{\cal A}_\psi (q;p)\right|^2 \ .
\label{expctnum4}
\end{eqnarray}

\section{Detector lifetime}

Integrals of the mode function products ${\cal A}_\chi (p;q)$ and ${\cal A}_\psi (q;p)$
appearing in Eqs.~(\ref{expctnum3}) and (\ref{expctnum4}) have the meaning of amplitudes for the detection processes $\psi\phi^*\to\chi$ and $\chi\phi\to\psi$, respectively.
These integrals
can be evaluated using the method developed in Refs.~\cite{Massar,Gabriel}.  Relegating details of the calculation to Appendices \ref{apx:amplitudepsi} and \ref{apx:amplitudechi}, here we only present the results.


For a $\psi$-particle detector, we are interested in the $\chi$-particle distribution in the final state, given by Eq.~(\ref{expctnum3}).  The amplitude ${\cal A}_\chi (p;q)$ appearing in that equation is given by 
\begin{equation}
|{\cal A}_\chi (p;q)| = \left({\pi\over \omega_p}\right)^{1/2} 
{ g|\beta_\phi\, \alpha_{\psi}^*| \over \sqrt{2eE}\  m_\chi} ~
 e^{{\pi\over 2}\sigma_+}
\left|\,W_{i\sigma_+,~i\sigma_-}\left(i\, \frac{\lambda_\chi}{2}\right)\,\right| \,. 
\label{ampli}
\end{equation}
Here, $\omega_p = \sqrt{p^2 + m_\chi^2}$ is the $\chi$-particle energy, $\beta_\phi$ is the negative-frequency Bogoliubov coefficient for $\phi$-particles, 
\beq
|\beta_\phi|^2 = \exp\left(-\frac{\pi m_\phi^2}{eE}\right) ,
\eeq
and $|\alpha_\psi|\approx 1$ is the positive-frequency Bogoliubov coefficient for $\psi$-particles.  Furthermore,
\begin{equation}
\sigma_{\pm} \equiv \frac{\lambda_\phi\pm\lambda_\psi}{4}\,,
\end{equation}
\beq
\lambda_i \equiv \frac{m_i^2}{eE}
\eeq
with $i=\phi,\psi,\chi$, and $W_{\sigma,\,\rho}\,(z)$ is the Whittaker function.  Note that the amplitude (\ref{ampli}) is independent of the $\psi$-particle momentum $q$.

This expression has a simple asymptotic form in the limit of large $m_\chi$, 
\beq
\lambda_\chi \gg \lambda_\psi \lambda_\phi.
\label{largemchi}
\eeq   
In this case,
\begin{equation}
 e^{{\pi\over 2}\sigma_+} 
\left|\,W_{i\sigma_+,~i\sigma_-}\left(i\, \frac{\lambda_\chi} {2}\right)\right| 
\approx 1  \,,
 \end{equation}
where we have used an asymptotic representation $W_{\lambda,\,\mu}(z)\approx e^{-z/2}z^\lambda$, which applies for $|z| \gg |\mu^2 - \lambda^2|$.  The $\chi$-particle distribution (\ref{expctnum3}) is then given by
\beq
dN_\chi =  \frac{g^2}{4eE m_\chi^2} \exp\left(-\frac{\pi m_\phi^2}{eE}\right) \frac{dp}{\omega_p}.
\label{dNchi}
\eeq
Note that the coupling $g$ in our model has dimension of mass squared, so the right-hand side of (\ref{dNchi}) is dimensionless, as it should be.

If in addition to (\ref{largemchi}) we assume that $m_\psi\gg m_\phi$, then the false detection processes of Figs.~1(a) and 1(c) are suppressed, and the $\chi$-particles in the final state count predominantly true detections.  As we shall see in a moment, the detection rate is proportional to $dN_\chi /dp$.  Eq.~(\ref{dNchi}) shows that this rate is proportional to the pair production rate of $\phi$-particles, as expected.



In the general case, the amplitude ${\cal A}_\chi (p;q)$ can be represented as
\begin{equation}
|\,{\cal A}_\chi (p;q)\,| = \omega_p^{-\frac{1}{2}}~ 
C_\chi \left(\,m_\chi,\,m_\psi,\,m_\phi\,\right).
\label{C}
\end{equation}
The corresponding $\chi$-particle distribution is
\begin{equation}
{dN_\chi} = {|C_\chi|^2 \over 2\pi} {dp \over \omega_p}. 
\label{dNchi1}
\end{equation}
The factor $dp / \omega_p$ on the right hand side can be recognized as the volume element of the Lorentz group.  This indicates that the distribution (\ref{dNchi1}) is invariant under Lorentz boosts.

 The final distribution (\ref{dNchi1}) can be represented as
\beq
\frac{dN_\chi}{dp} = \int d\tau\,\frac{dN_\chi}{dp\,d\tau}\,,
\label{dNchidp}
\eeq
where ${dN_\chi}/{dp\,d\tau}$ is the expectation number of $\chi$-particles emitted in the momentum interval $dp$ per unit proper time $\tau$ along the detector trajectory.  The latter quantity transforms under Lorentz boosts as $\omega_p^{-1}$; hence we can write 
\beq
\frac{dN_\chi}{dp\,d\tau} = \frac{1}{\omega_p} f(I)\,,
\label{Nf}
\eeq
where the function $f$ is Lorentz invariant and can therefore depend only on the invariant
\beq
I = \epsilon_{\mu\nu}\,p^\mu q_{\rm phys}^\nu(\tau)\,.
\eeq
(The other invariant, $I' = p_\mu q_{\rm phys}^\mu(\tau)$ is related to $I$ by ${I'}^2 = I^2 + m_\psi^2 m_\chi^2$.)
The physical momentum $q_{\rm phys}(\tau)$ specifies the rest frame of the detector at the time of measurement\footnote{Here we have assumed that the distribution (\ref{Nf}) depends only on the local value of the detector momentum $q_{\rm phys}(\tau)$.  This is justified, as long as $q_{\rm phys}$ remains essentially unchanged during the time of interaction.}. Note that in this frame the invariant $I$ is simply proportional to the $\chi$-particle momentum, $I = -m_\psi p$.

For a hyperbolic detector trajectory, we have
\beq
d\tau = \frac{1}{a_\psi}\frac{dq_{\rm phys}}{\omega_{q,\rm phys}}\,,
\eeq
where $a_\psi = eE/m_\psi$ is the proper acceleration and $\omega_{q,\rm phys} = \sqrt{q_{\rm phys}^2 + m_\psi^2}$ is the detector energy.  Substituting this in (\ref{dNchidp}) and comparing with (\ref{dNchi1}), we can identify
\beq
\frac{|C_\chi |^2}{2\pi} = \frac{1}{a_\psi} \int \frac{dq_{\rm phys}}{\omega_{q,\rm phys}} f(I)\,.
\label{Cchi1}
\eeq
On the other hand, it follows from Eq.~(\ref{Nf}) that
\beq
\frac{dN_\chi}{d\tau} = \int \frac{dp}{\omega_p} f(I)\,,
\label{dNchidtau}
\eeq
and it is easily understood that the integrals over the Lorentz group in the last two equations are equal to one another.  Hence, the scattering rate in the detector's frame can be expressed as
\begin{equation}
{dN_\chi \over d\tau} = {|C_\chi |^2\over 2\pi}  {eE \over m_\psi}\,.
\end{equation}

The characteristic collision time is
\beq
\tau_\psi = \frac{2\pi m_\psi}{|C_\chi|^2 eE}\,.
\label{taupsi}
\eeq
This is the average proper time that it takes for a detector to experience a collision.  It can be called the detector lifetime.  As one could expect, this lifetime is independent of the rest frame of the detector: it is the same for all points on the hyperbolic detector trajectory.  


For a $\chi$-particle probe, the detection process is $\chi\phi\to\psi$.  
The amplitude ${\cal A}_\psi$ appearing in Eq.~(\ref{expctnum4}) is evaluated in Appendix~\ref{apx:amplitudechi}.
The result is rather cumbersome, and we do not reproduce it here.
As before, the amplitude is independent of $q$ and can be represented as
\begin{equation}
|\,{\cal A}_\psi (q;p)\,| = \omega_p^{-\frac{1}{2}}~ 
C_\psi \left(\,m_\chi,\,m_\psi,\,m_\phi\,\right)\,.
\label{C1}
\end{equation}
The corresponding detector lifetime is
\beq
\tau_\chi = \frac{2\pi m_\chi}{|C_\psi|^2 eE}\,.
\eeq
This is the same as $\tau_\psi$ in Eq.~(\ref{taupsi}) with the replacement $m_\psi\to m_\chi$ and $C_\chi \to C_\psi$.  Once again, we see that the detection rate is independent of the detector's rest frame.

\section{Momentum distribution of the pairs}

Our results in Section 6 indicate that the detection rate of $\phi\phi^*$-pairs is independent of the state of motion of the detector.  
This is in agreement with the Lorentz invariance of the in-vacuum state that we established in Sec.~3.

However, the analysis of Sec.~6 does not allow us to determine the momentum distribution of the pairs measured by the detector.  Indeed, it is clear from that analysis that it can only determine the distribution integrated over momenta or (which is essentially the same) over the proper time along the detector trajectory.
In particular, we do not have an answer to the question we asked in the introduction: are the $\phi$ and $\phi^*$ particles of the pairs observed to nucleate at rest and then move away from one another, or they can also be detected on the `wrong' parts of their hyperbolic trajectories, where they move towards one another at a high speed?  In the former case, the distribution should be cut off at $k_{\rm phys}\approx 0$, as in Eq.~(\ref{theta}).

In order to address this question, we consider a detector with a time-dependent coupling
\beq
g(t) = g\, e^{-t^2/T^2}\,,
\label{gt}
\eeq
so the detector is effectively turned on only for a finite time interval $\Delta t \sim T$ around $t=0$.
Focusing on a charged $\psi$-detector, we shall assume that
\beq
m_\phi \ll 
eET \ll m_\psi\,,
\label{eET}
\eeq
so that the detector momentum $q_{\rm phys}$ remains essentially unchanged during this interval (while $\phi$-particles can be significantly accelerated). 
 For example, if we choose $q=0$, then $q_{\rm phys}\approx 0$ during the interaction.  The resulting $\chi$-particle will then have (approximately) the same momentum as the $\phi^*$-particle that collided with $\psi$,
\begin{equation}
p\approx k_{\rm phys}\,.
\end{equation}
 We assume as before that 
\beq
m_\phi \ll m_\psi < m_\chi - m_\phi\,.  
\label{ineq1}
\eeq
In addition, we may assume that turning on and off of the detector is adiabatic, 
\beq
T\gg (p^2 + m_\phi^2)^{-1/2}\,,
\label{Tm}
\eeq
so that the time variation of the coupling does not lead to a sizable violation of 
energy conservation at the moment of collision. In fact, the first strong inequality in 
(\ref{eET}) implies
\begin{equation}
T \gg \frac{\lambda_\phi}{m_\phi}\,, \label{Tlp}
\end{equation}
so (\ref{Tm}) is automatically satisfied in the regime where pair nucleation is rare, $\lambda_\phi = m^2_\phi/(eE) \gg 1$,
in which we are now focusing.  It then also follows from Eqs.~(\ref{eET}) and (\ref{Tlp}) that 
\beq
eET^2 \gg \lambda_\phi \gg 1\,.   \label{eET2}
\eeq

To find the distribution of $\chi$-particles, we need to calculate the amplitude
\beq
{\cal A}_\chi(p;q=0) = \int_{-\infty}^\infty dt~ g(t)~ \phi^*_{-p}(t)\,\psi_0(t)\,\chi^*_p(t)\,. 
\label{Agt}
\eeq
Since the $\psi$-particle is practically at rest during the time interval of interest, we can use
\beq
\psi_0(t) \approx \frac{1}{\sqrt{2m_\psi}}\, e^{-im_\psi t}\,.
\label{psi0}
\eeq
The amplitude (\ref{Agt}) is calculated in Appendix \ref{apx:timedependent}.  Here we are only interested in the $p$-dependence of the amplitude.  
In the parameter regime specified above, this is given by the factor
\beq
\frac{dN_\chi}{dp}  \propto  |{\cal A}_\chi(p;q=0)|^2 \propto \exp \left[-\frac{2\,(p+\omega_p -m_\psi)^2}{(eET)^2}\right]\,.
\eeq
This distribution is peaked at $p={\bar p}$, which is specified by 
\beq
{\bar p} +\sqrt{{\bar p}^2 + m_\chi^2} = m_\psi\,, 
\eeq
or
\beq
{\bar p}=-\frac{1}{2m_\psi} (m_\chi^2 - m_\psi^2)
\label{barp}
\eeq
and has width 
\begin{equation}
\Delta p \sim eET\,. \label{widthp}  
\end{equation}
 Note that we will only see a sharp peak in the distribution provided that $|\bar p| \gg \Delta p \sim eE T$. Using (\ref{eET}), this requires
\begin{equation}
\bar p^2 \gg m_\phi^2\,. \label{rel}
\end{equation}
Eq.~(\ref{barp}) is to be compared with Eq.~(\ref{Mmm}) for the kinematically allowed values of $p$.  For $q=0$, it gives 
\begin{equation}
p\approx \pm \frac{1}{2m_\psi} (m_\chi^2 - m_\psi^2)\,.  \label{kin}
\end{equation}
In deriving (\ref{kin}), we have assumed $m_\chi^2 - m_\psi^2 \gg m_\phi m_\psi$, which in turn implies the ultrarelativistic motion of the $\phi$ particle, as in (\ref{rel}).

From (\ref{barp}), it is clear that only the negative value in (\ref{kin}) is actually detected. This is consistent with the picture of pairs being produced preferentially at rest in the observer's frame, and then being accelerated by the electric field. In this
case, $\phi$-antiparticles move in the direction opposite to $E$ and hit
the $\psi$-probe with $k_{\rm phys} < 0$. This
is imprinted in the momentum of the final $\chi$-particle, $p \approx k_{\rm phys}
< 0$. If pairs were to nucleate at arbitrarily large speed with respect to
the detector, then the detector would also experience collisions with
antiparticles moving in the direction of $E$, in contradiction with the
fact that we do not find a peak in the distribution of detected particles
for the positive value in (\ref{kin}).

With our assumptions, we find from (\ref{rel}) that the asymmetry in the momentum distribution is visible only for relativistic particles. In this case, the precision to which we can determine the frame of nucleation is limited by the acceleration time 
\begin{eqnarray}
\Delta t_{\rm acc} \sim \frac{m_\phi}{eE}\,,   \label{instsize}
\end{eqnarray}
 (this is comparable to the size of the instanton). Nonetheless, it seems in principle possible to increase this precision by relaxing the first inequality in (\ref{eET}), or by using a different type of model detector. This possibility is currently under investigation \cite{GKT}.

\section{Conclusions}

Our results indicate that all inertial observers in a $(1+1)$-dimensional Minkowski universe with a constant electric field detect the same distribution (\ref{theta}), (\ref{thetaanti}) of charged particle pairs created by the Schwinger process.  In each observer's frame, particles and antiparticles nucleate preferentially at rest and are then accelerated in opposite directions by the electric field. 
This implies that particles that have already been formed and are ready to interact with a particular detector, may not yet exist from the point of view of another
detector which moves relative to the first\footnote{This scenario was suggested by 
N. Tanji \cite{Tanji:2008ku}, although he did not provide evidence to support it.}.
This picture fits well with the Lorentz invariance of the in-vacuum state which we established in Section 3. 

The situation here is similar to that for the Bunch-Davis vacuum in de Sitter space.  This state is de Sitter invariant, but each geodesic observer sees a thermal bath in her rest frame
with the same universal temperature $T=H/(2\pi)$
where $H$ is the inverse of the curvature radius of the de Sitter space.

The close analogy between pair creation in $(1+1)D$ and bubble nucleation leads us to expect that a similar picture should apply to observation of bubble nucleation (in (3+1) as well as in other dimensions).  An observer ${\cal O}$ who fills the space with detectors at rest with respect to himself will see bubbles nucleating at minimal radius at rest and then expanding in the outward direction.  Another observer ${\cal O}'$, who has installed detectors at rest in her frame, sees the same picture.  But if ${\cal O}'$ were to watch the detectors of ${\cal O}$, she would see a very different scenario.  A small piece of the bubble wall first appears and moves at a very high speed towards the bubble center.  It gradually slows down and grows in size, and eventually bounces at the minimum radius when the wall extends over half a sphere.  After the bounce, the wall expands, and still later it closes up to a full sphere. 

We also discussed the unusual, and to a certain extent unphysical, properties of the in-vacuum state, which put the above stated conclusions somewhat in doubt.  The singularity structure of the two-point 
function in the in-vacuum state does not have Hadamard form, and as a result the expectation values of physical observables cannot be regulated in a Lorentz invariant way.  The same could apply to 
regularization of the loop diagrams in higher orders of perturbation theory.
This 
violation of Lorentz invariance has no effect on the tree-level scattering processes that we discussed in this paper.  However, it does raise some questions about the overall consistency of the model.

In a more realistic model, the electric field would have to be switched 
on at some initial time in the past. In order to investigate the effect of 
initial conditions, one would have to do similar calculations for an 
electric field which is turned on for a finite period of time $T$ and 
check whether or not the results approach our results in this paper in 
the limit $T\to\infty$.  It would also be interesting to extend the 
analysis to de Sitter space\footnote{Schwinger pair production in  (1+1)-dimensional de Sitter space has been considered in Ref. \cite{Garriga}.}.  We hope to address these problems in future work.

\acknowledgments
We are grateful to Takahiro Tanaka for very useful discussions.  A.V. also wants to thank Slava Mukhanov for many stimulating discussions of the possible observer dependence of bubble nucleation over the years.  This work was supported in part 
by grant PHY-0855447 from the National Science 
Foundation, the JSPS Grants-in-Aid for Scientific Research (C) No.~22540274 and (A) No.21244033, No.22244030, the Grant-in-Aid for Creative Scientific Research No.~19GS0219, the Grant-in-Aid for Scientific Research on Innovative Area No.21111006, AGAUR 2009-SGR-168, MEC FPA 2010-20807-C02- 02 and CPAN CSD2007-00042 Consolider-Ingenio 2010. J.G. thanks the Tufts Cosmology Institute for hospitality during the preparation of this work.

\appendix

\section{Klein-Gordon product}
\label{apx:KGproduct}

Here we calculate the Klein-Gordon product of mode functions, defined in 
Eq.~(\ref{KGdef}). Using the integral representation (\ref{Dintrep}) for the parabolic cylinder functions, we can write,
up to normalization, 
\begin{equation}
\phi_k(t,x) =  e^{i \frac{z^2}{2}}
\int_0^{\infty} d w\ e^{(1-i) z w - \frac{w^2}{2}}\  w^{\nu}\,,
\label{Ukint}
\end{equation}
and
\begin{equation}
\bar\phi_{\bar k}(t,x) = 
 e^{-ie \Omega}\, e^{i \frac{\bar z^2}{2}}
\int_0^{\infty} d w' \ e^{(1-i) \bar z w' - \frac{w'^2}{2}}\  w'^{\nu}.
\end{equation}
Substituting in (\ref{KGdef}) we have
\begin{equation}
K=\int_0^\infty dw\   w^{\nu} e^{-\frac{w^2}{2}} 
\int_0^{\infty} dw'\ w'^{\nu} e^{- \frac{w'^2}{2}} I(k,\bar k, w, w',t),
\end{equation}
where
\begin{equation}
I(k,\bar k, w, w',t) =  i\ (eE)^{1/2} \int_{t = {\rm const}} dx\, 
[(1-v) \gamma \bar z - z- (1+i) (\gamma w'-w)] f_k\bar f_{\bar k}
\label{intdx}
\end{equation}
and
\begin{equation}
f_k = e^{-ik x}\, e^{i\frac{z^2}{2}}\, e^{(1-i) z w}\,, \quad \bar f_{\bar k} = 
e^{i\bar k \bar x-i e \Omega}\, e^{i\frac{\bar z^2}{2}}\, e^{(1-i) \bar z w'}.
\end{equation}  

Let us consider the mixing of the positive with the negative 
frequency modes. 
For convenience, we choose the integration surface at $t=0$ and take $k=0$. In this 
case, Eqs.~(\ref{barz}) give $\bar x=\gamma x$, $\bar z= -\sqrt{eE} \gamma v x +\bar k/\sqrt{eE}$,
 and  we have
\begin{equation}
I =  {i\over \gamma v} e^{i {\bar k^2\over 2veE}}\int_{t = 0} 
d\bar z \,  [(1-v) \gamma \bar z - (1+i) (\gamma w'-w)]\,
e^{-i {(1-v)\over 2v} \bar z^2+(1-i)\bar z w'} \,.
\label{intdz}
\end{equation}
Performing the Gaussian integral, we have
\begin{equation}
I =  {2 \sqrt{\pi} i\over \gamma \sqrt{(1-v) v}}\,
 e^{i {\bar k^2\over 2veE}-{v w'^2\over (1-v) }}\,( w-\gamma (1+v) w')\,.
\label{I}
\end{equation}
Finally, performing the $w$ and $w'$ integrals, we obtain
\begin{equation}
K=0\,.
\end{equation}


\section{Two-point function}
\label{apx:2pfunction}

In this Appendix we shall calculate the gauge-invariant 2-point functions $G^\pm$, defined by Eqs.~(\ref{G+}), (\ref{G-}).  We consider the in-vacuum expectation value,
\begin{equation}
G^+(x^\mu,y^\mu) \equiv {}_{\rm in}\langle\,0\,|\,
\phi^\dagger(x^\mu)~ e^{-ie \int_{x}^{y} A_\nu dx^\nu} ~\phi(y^\mu)\,
|\,0\,\rangle_{\rm in}\,.
\label{G+A}
\end{equation}
Using the expression for the $\phi$-field 
given by Eq.~(\ref{phi-field}), we find
\begin{eqnarray}
2 \pi G^+&=&e^{-ie \int A_\nu dx^\nu} 
\int dk\  \phi_{-k} (t)\, \phi^*_{-k}(t')\, e^{ik(x-y)}
\cr
&=&e^{-ie \int A_\nu dx^\nu}  
\int dk\  \phi_{k} (t)\, \phi^*_{k}(t')\, e^{ik(y-x)}\,.
\end{eqnarray}
Using the integral representation of $\phi_k$ from (\ref{Dintrep}),
 the integral over momenta leads to a delta function, and we have
\begin{eqnarray}
G^+&=& \frac{|\alpha|^2}{4\pi} e^{-ie \int A_\nu dx^\nu} 
\int_0^\infty dX~X^\nu e^{\frac{i}{2}X^2}
\int_0^\infty dX'~ X'^{\nu^*} e^{-\frac{i}{2}X'^2} 
\cr&&
\times e^{\frac{i}{2}eE (t^2-t'^2)-i \sqrt{2eE}\, (tX-t'X')} 
\delta\left(X'-X-S_-\right)\,,
\end{eqnarray}
where $\alpha$ is the Bogoliubov coefficient given by
Eq.~(\ref{bogoliubov}). 
We also define
\begin{equation}
S_\pm \equiv \sqrt{\frac{eE}{2}}~ (\Delta t \pm \Delta x)\,,
\qquad \Delta t\equiv t'-t\,,\qquad \Delta x \equiv y-x
\end{equation}

Assuming $S_- >0$, we have $X'>X$. Doing the $X'$ integral we obtain
\begin{equation}
G^+= {|\alpha|^2 \over 4\pi} e^{-ie \int  A_\nu dx^\nu}
\int dX X^{\nu} (X+S_-)^{\nu^*} 
e^{-\frac{i}{2}S_-^2
-iXS_-+\frac{i}{2}eE(t^2-t'^2)- i\sqrt{2eE}\,[(t-t')X-t' S_-]} \,.
\label{pve}
\end{equation}
This can be rewritten as
\begin{equation}
G^+={|\alpha|^2\over 4\pi} e^{-ie \int  A_\nu dx^\nu}
 e^{\frac{i}{2}eE(t^2-t'^2) + i\sqrt{2eE}\, t' S_- -\frac{i}{2} S_-^2} 
\int dX (X+S_-)^{\nu^*}  X^{\nu}e^{iS_+ X}\,,
\end{equation}
and, after some rearrangements,
\begin{equation}
G^+={|\alpha|^2\over 4\pi} e^{-ie \int  A_\nu dx^\nu}
e^{-\frac{i}{2}eE \Delta x (t+t')}
(-i S_+S_-)^{-\frac{1}{2}}~ \Gamma(-\nu^*)~ 
W_{i\lambda/2\,,\,0}\,(-i S_+S_-)\,,
\end{equation} 
where $W_{\sigma,\,\rho}\,(z)$ is the Whittaker function.

In the gauge~(\ref{gauge}), if we integrate along a straight line, 
so that $dx = (\Delta x/\Delta t) dt$, 
then the Wilson line is 
\begin{equation}
\int  A_\nu dx^\nu = -{ E \over 2} \Delta x (t+t'),
\end{equation}
so that the two-point function is given by:
\begin{equation}
G^+(\Delta x^\mu)=\frac{|\alpha|^2}{4\pi}
 \Gamma(-\nu^*)\,
 \frac{W_{i\lambda/2,\,0}\left(i eE\,\Delta s^2/2\right)}
{\sqrt{i eE\,\Delta s^2/2}}\,, 
\qquad {\rm for} \qquad
\Delta t-\Delta x >0 
\label{60}
\end{equation}
where $\Delta s^2= -(\Delta t)^2+(\Delta x)^2$. 

A similar 
calculation for $S_- < 0$ gives the complex conjugate expression:
\begin{equation}
G^+(\Delta x^\mu)=\frac{|\alpha|^2}{4\pi}
 \Gamma(-\nu)\,
\frac{W_{-i\lambda/2,\,0}\left(-i eE\,\Delta s^2/2\right)}
{\sqrt{-i eE\,\Delta s^2/2}}\,,
\qquad {\rm for} \qquad
\Delta t-\Delta x < 0 
\label{61}
\end{equation}

Likewise, we may consider the 2-point function
\begin{equation}
G^-(x^\mu,y^\mu)= {}_{\rm in}\langle\,0\,|\,
\phi(y^\mu)~ e^{-ie \int_{x}^{y} A_\nu dx^\nu}~\phi^\dagger (x^\mu)\,
|\,0\,\rangle_{\rm in}\,,
\label{G-A}
\end{equation}
which can be expressed in terms of the mode functions as 
\begin{equation}
2\pi G^-(\Delta t,\Delta x)=e^{-ie \int A_\nu dx^\nu}  
\int dk~  \phi_{k} (t')\,\phi^*_{k}(t)\,e^{ik(y-x)}\,. 
\end{equation}
Comparing this with Eq.~(\ref{G+A}), we see that it is simply related to $G^+$,
\begin{equation}
G^-(\Delta t,\Delta x)=G^+(-\Delta t,\Delta x)\,.
\end{equation}

\section{Lorentz violating current}
\label{apx:LVcurrent}

In this Appendix, we evaluate the expectation value of the current $J_\mu$ in the in-vacuum, using a direct momentum space regularization.  

Substituting the mode expansion (\ref{phi-field}) in Eq.~(\ref{currentform}), we have $\langle J_0 \rangle = 0$ and
\beq
\left\langle J_1\right\rangle = 2e\int \frac{dk}{2\pi}(k+eEt) \left|\phi_k\right|^2 .
\label{J1}
\eeq
Here, $\phi_k$ are functions of $z = (eE)^{-1/2}(k+eEt)$, so the integrand in Eq.~(\ref{J1}) is a function only of $z$.  It appears that we can perform a change of variable $k\to k-eEt$
to make the integral independent of $t$.  However, such a shift of integration variable is not always legitimate for divergent integrals.  

Let us consider
\beq
\left\langle\frac{\partial J_1}{\partial t}\right\rangle = 2e^2 E \int \frac{dk}{2\pi} \frac{d}{dk} \left[(k+eEt) |\phi_k|^2\right]  = \frac{e^2 E}{\pi} 
\left[(k+eEt) |\phi_k|^2\right]_{k\to -\infty}^{k\to +\infty}.
\eeq
This shows that $\langle{\partial j_1}/{\partial t}\rangle$ is nonzero if the function
\beq
F(k,t) \equiv (k+eEt) \left|\phi_k\right|^2
\label{Fk}
\eeq
has different limits at $k\to \pm\infty$.

Now, the asymptotic forms of $\phi_k(z)$ are given by
\beq
\phi_k(z) \approx f_k(z)\,, \qquad {\rm for}\quad z\to -\infty
\label{u-}
\eeq
and
\beq
\phi_k(z) \approx \alpha f_k^*(z) + \beta f_k(z)\,, \qquad {\rm for}\quad z\to -\infty,
\label{u+}
\eeq
where
\beq
f_k(z) \approx  (eE)^{-\frac{1}{4}}~{e^{i\frac{z^2}{2}}\over (\sqrt{2}|z|)^{1-i\lambda\over 2}}.
\label{fk}
\eeq
Substituting this in (\ref{Fk}), we have
\beq
F(k\to -\infty) = \frac{1}{2}.
\eeq
For $k\to +\infty$, $F(k)$ does not approach a definite limit, as it includes rapidly oscillating terms:
\beq
F(k\to +\infty) = \frac{1}{2}\left(|\alpha|^2 + |\beta|^2\right) + ({\rm oscillating ~ terms}).
\eeq

The oscillating terms indicate that $\partial J_1/\partial t$ is not a well defined quantity in the in-vacuum.  In order to remove these terms, let us consider 
\beq
\left\langle J_1(t)-J_1(t')\right\rangle = \int_{t'}^t d\tau \frac{\partial j_1}{\partial \tau}(\tau) = \frac{e^2 E}{\pi}  \int_{t'}^t d\tau 
\left[(k+eE\tau) |u_k|^2\right]_{k\to -\infty}^{k\to +\infty}.
\label{C2}
\eeq
We can evaluate this by first integrating over $\tau$ and then taking the limits $k\to\pm\infty$.
This corresponds to the usual subtraction procedure, where we first do the subtraction in the integrand and then integrate over momenta.

The oscillating term is proportional to
\beq
e^{iz^2} \propto e^{2ikt}.
\eeq
This oscillates infinitely fast as $k\to\infty$, so the oscillating term in (\ref{C2}) vanishes in the limit, and we obtain
\beq
\left\langle J_1(t)-J_1(t')\right\rangle =  \frac{e^2 E}{\pi} e^{-\frac{\pi m^2}{eE}} (t-t').
\label{3}
\eeq
Here, we have used
\beq
|\alpha|^2 = 1 + |\beta|^2
\eeq
and
\begin{equation}
|\beta|^2 = e^{-\frac{\pi m^2}{eE}}.
\label{bk}
\end{equation}
Finally, taking the limit $t'\to t$, we obtain
\beq
\left\langle \frac{\partial J_1}{\partial t}\right\rangle =  \frac{e^2 E}{\pi} e^{-\frac{\pi m^2}{eE}}.
\eeq

\section{Transition amplitude for a $\psi$-detector}
\label{apx:amplitudepsi}

Here we shall calculate the integral over mode functions appearing in Eq.~(\ref{expctnum3}) for ${\cal A}_\chi(p;q)$ following the method developed in Ref.~\cite{Gabriel}. 

We first express the mode functions (\ref{sol:in}) for $\phi$ and $\psi$ fields using the integral representation (\ref{Dintrep}) of parabolic cylinder functions. 
The neutral $\chi$-particle mode function is simply given by a plane wave. Thus we have 
\begin{eqnarray}
\phi_k&=&\frac{\alpha_\phi~ e^{-i\frac{\pi}{4}}}{(2\pi)^{1/2} (2eE)^{1/4}}~
e^{\frac{i}{2}z_\phi^2} \int_{0}^{\infty}dX~ X^{\nu_\phi}~ 
e^{-i\sqrt{2}z_\phi\, X\,
+\, i\frac{X^2}{2}}\,,
\cr
\psi_q&=&\frac{\alpha_\psi~ e^{-i\frac{\pi}{4}}}{(2\pi)^{1/2} (2eE)^{1/4}}~
e^{\frac{i}{2}z_\psi^2}\int_{0}^{\infty}dY~ Y^{\nu_\psi}~
e^{-i\sqrt{2} z_\psi\, Y\,
+\, i\frac{Y^2}{2}}\,,
\cr
\chi_p&=&\frac{1}{(2\omega_p)^{1/2}}e^{-i\omega_p t}\,,
\label{solutions}
\end{eqnarray}
where the Bogoliubov coefficient $\alpha$ defined in (\ref{bogoliubov}) is used with an index of each particle instead of its momentum, because we found that there was no dependence on momentum. 
We have also defined
\begin{eqnarray}
&&z_\phi = \sqrt{eE}\left(t+\frac{k}{eE}\right)\,,
\qquad
\nu_\phi = -\frac{1+i\lambda_\phi}{2}\,,
\qquad
\lambda_\phi = \frac{m_\phi^2}{eE}\,,
\cr
&&z_\psi = \sqrt{eE}\left(t+\frac{q}{eE}\right)\,,
\qquad
\nu_\psi = -\frac{1+i\lambda_\psi}{2}\,,
\qquad
\lambda_\psi = \frac{m_\psi^2}{eE}\,.
\label{variables}
\end{eqnarray}

Plugging Eqs.(\ref{solutions}) and (\ref{variables}) into (\ref{expctnum3}),  the amplitude
${\cal A}_\chi(p;q)$ becomes
\begin{eqnarray}
{\cal A}_\chi (p;q) 
&=& 
\frac{g\,\alpha_{\phi}^*~ \alpha_\psi}{(2eE)^{1/2}(2 w_p)^{1/2}}~ e^{\frac{i}{2eE}(2qp-p^2) }
 \int_0^\infty dX~ X^{\nu_\phi^*}~ e^{i\sqrt{\frac{2}{eE}}(q-p)X\, - \, i\frac{X^2}{2}}
\nonumber\\
&&\times 
\int_0^\infty dY~ Y^{\nu_\psi}~ e^{-i\sqrt{\frac{2}{eE}}\, q Y \, + \, i\frac{Y^2}{2}} 
~\frac{1}{2\pi}\int^\infty_{-\infty} dt~ e^{i(p+\omega_p + \sqrt{2eE} (X-Y))t}\ .
\end{eqnarray}
The time integration gives rise to the delta function
\begin{equation}
\delta\left(\sqrt{2eE} (X-Y)  + p +\omega_p\right).
\end{equation}
The delta function tells us that $Y>X$, and so we can perform the $Y$ integration first. 
Then, we have
\begin{eqnarray}
{\cal A}_\chi (p;q) = 
\frac{g\,\alpha_{\phi}^*~ \alpha_\psi}{2eE\,(2 \omega_p)^{1/2}}\,
e^{\frac{i}{2eE}(2qp-p^2) -i\sqrt{\frac{2}{eE}}\,q \mu_+\, + \, i \frac{\mu_+^2}{2}}
\int_0^\infty dX\, X^{\nu_\phi^*}\,
 (X+\mu_+)^{\nu_\psi}~
 e^{i \mu_- X }\,,
\end{eqnarray}
where
\begin{equation}
\mu_\pm \equiv {\omega_p \pm p \over \sqrt{2eE}}.
\end{equation}
The remaining integral can be expressed in terms of the Whittaker function.
By using  
\begin{equation}
\mu_+\mu_- = {m_\chi^2\over 2eE} \equiv \frac{\lambda_\chi}{2}\,,
\end{equation}
we finally obtain
\begin{equation}
|{\cal A}_\chi (p;q)| = \left({\pi\over \omega_p}\right)^{1/2} 
{g |\beta_\phi\, \alpha_{\psi}^*| \over \sqrt{2eE}\  m_\chi} ~
 e^{{\pi\over 2}\sigma_+}
\left|\,W_{i\sigma_+,~i\sigma_-}\left(i\, \frac{\lambda_\chi}{2}\right)\,\right| \,, \label{ampli}
\end{equation}
where
\begin{equation}
\sigma_{\pm} \equiv \frac{\lambda_\phi\pm\lambda_\psi}{4}\,
\end{equation}
and $|\beta_\phi|\equiv e^{-{\pi \lambda/2}}$ is the negative frequency 
Bogoliubov coefficient.

\section{Transition amplitude for a $\chi$-detector}
\label{apx:amplitudechi}

Now we calculate the transition amplitude ${\cal A}_\psi(q;p)$ in (\ref{expctnum4}).

Using the identity
\begin{eqnarray}
D_{\nu^*}[-(1-i)z] = \frac{\Gamma (\nu^*+1)}{\sqrt{2\pi}} \left[~
e^{\frac{i}{2}\pi \nu^*} D_{-\nu^*-1}[-(1+i)z] 
+ e^{-\frac{i}{2}\pi \nu^*} D_{-\nu^*-1}[(1+i)z]
~\right]\,
\label{identity}
\end{eqnarray}
and the integral representations (\ref{solutions}), we have, up to an overall phase factor, 
\begin{eqnarray}
 {\cal A}_\psi (q;p) &=& g \int_{-\infty}^{\infty}dt~\phi_{q-p}^*~\psi_q^*~\chi_p
\cr
&=&\frac{g\,\alpha_\psi^*}{(2eE)^{1/2}(2\omega_p)^{1/2}}~
e^{-i \pi\nu_\phi}~ e^{\frac{i}{2eE}(p^2-2qp) }
\int_0^\infty dX~X^{\nu_\phi}~e^{-i\sqrt{\frac{2}{eE}}(q-p)X\, + \, i\frac{X^2}{2}}
\cr
&&\quad\times 
\int_0^\infty dY~ Y^{\nu_\psi^*}~ e^{i\sqrt{\frac{2}{eE}}\, q Y \, - \, i\frac{Y^2}{2}} 
~\frac{1}{2\pi}\int^\infty_{-\infty} dt~ e^{i(-p-\omega_p - \sqrt{2eE} (X-Y))t}
\cr
&&+\frac{g\,\alpha_\psi^*}{(2eE)^{1/2}(2\omega_p)^{1/2}}~
e^{\frac{i}{2eE}(p^2-2qp) }
\int_0^\infty dX~X^{\nu_\phi}~e^{i\sqrt{\frac{2}{eE}}(q-p)X\, + \, i\frac{X^2}{2}}
\cr
&&\quad\times 
\int_0^\infty dY~ Y^{\nu_\psi^*}~ e^{i\sqrt{\frac{2}{eE}}\, q Y \, - \, i\frac{Y^2}{2}} 
~\frac{1}{2\pi}\int^\infty_{-\infty} dt~ e^{i(-p-\omega_p + \sqrt{2eE} (X+Y))t}
\ .
\label{calopsi}
\end{eqnarray}
The time integration in the first term in Eq.~(\ref{calopsi}) give a delta function
\begin{eqnarray}
\delta\left(-p-\omega_p-\sqrt{2eE} (X-Y) \right)\,.
\end{eqnarray}
This delta function tells us that $Y>X$, so we can perform the $Y$ integration first. 
Then the first term is
\begin{eqnarray}
\frac{g\,\alpha_{\psi}^*}{2eE\,(2 \omega_p)^{1/2}}\,
e^{-i \pi\nu_\phi}\,
e^{\frac{i}{2eE}(p^2-2qp) +i\sqrt{\frac{2}{eE}}\,q \mu_+\, - \, i \frac{\mu_+^2}{2}}
\int_0^\infty dX\, X^{\nu_\phi}\,
 (X+\mu_+)^{\nu_\psi^*}~
 e^{-i \mu_- X }\,,
\label{1}
\end{eqnarray}
where
\begin{equation}
\mu_\pm \equiv {\omega_p \pm p \over \sqrt{2eE}}.
\end{equation}

On the other hand, the time integration in the second term of (\ref{calopsi}) gives  
\begin{eqnarray}
\delta\left(-p-\omega_p+\sqrt{2eE} (X+Y)\right)\,.
\end{eqnarray}
Performing the $Y$-integration first, we find
\begin{eqnarray}
\frac{g\,\alpha_{\psi}^*}{2eE\,(2 \omega_p)^{1/2}}\,
e^{\frac{i}{2eE}(p^2-2qp) +i\sqrt{\frac{2}{eE}}\,q \mu_+\, - \, i \frac{\mu_+^2}{2}}
\int_0^{\mu_+} dX\, X^{\nu_\phi}\,
 (-X+\mu_+)^{\nu_\psi^*}~
 e^{i \mu_- X }\,.
\label{2}
\end{eqnarray}
The remaining integrals of Eqs.(\ref{1}) and (\ref{2}) can be expressed in terms of Whittaker functions.
Using  
\begin{equation}
\mu_+\mu_- = {m_\chi^2\over 2eE} \equiv \frac{\lambda_\chi}{2}\,,
\end{equation}
we finally obtain
\begin{eqnarray}
|{\cal A}_\psi (q;p)| &=& \left({\pi\over \omega_p}\right)^{1/2} 
\left|{\alpha_{\psi}^*\,\beta_\phi \over \alpha_\phi}\right|~
{ g \over \sqrt{2eE}\  m_\chi} ~
 e^{{\pi\over 2}\sigma_+}
\cr
&&\times
\left|\,\beta_\phi^*~W_{i\sigma_+,~i\sigma_-}\left(i\, \frac{\lambda_\chi}{2}\right)\,
+
\frac{\Gamma(-\nu_\psi)}{\Gamma(-\nu^*_{\phi}-\nu_\psi)}~
M_{i\sigma_+,~i\sigma_-}\left(i\, \frac{\lambda_\chi}{2}\right)\,
\right| \,, 
\label{ampli1}
\end{eqnarray}
where
\begin{equation}
\sigma_{\pm} \equiv \frac{\lambda_\phi\pm\lambda_\psi}{4}\,.
\end{equation}

\section{Amplitudes for a time-dependent coupling}
\label{apx:timedependent}

To evaluate the amplitude ${\cal A}_\chi(p;q=0)$ in Eq.~(\ref{Agt}), we first substitute the plane wave solutions for $\psi_0$ and $\chi^*_p$ and the integral representation for $\phi_{-p}^*$,
\begin{eqnarray}
{\cal A}_\chi(p;q=0) &=& \int_{-\infty}^\infty dt~ g(t)~ \phi^*_{-p}(t)\,\psi_0(t)\, \chi^*_p(t)
\cr
&=& 
C\,\omega_p^{-1/2} \int_0^\infty dX\, X^{\nu_\phi^*}\, e^{-i\frac{X^2}{2}}  \int_{-\infty}^\infty dt\, e^{-\frac{t^2}{T^2}}\,e^{-i\frac{z^2}{2}}\, e^{i\sqrt{2} zX},
\end{eqnarray}
where
\beq
z=\sqrt{eE} \left(t-\frac{p}{eE}\right)
\eeq
and
\beq
C=\frac{g\,\alpha_\phi \,e^{i\pi/4}}{\sqrt{8\pi m_\psi}(2eE)^{1/4}}
\eeq
is a constant.  After completing the square in the exponent of the $t$-integral and performing the Gaussian integration, this becomes
\beq
{\cal A}_\chi(p;q=0) = CT \omega_p^{-1/2} \sqrt{\frac{2\pi}{2+ieET^2}}\,e^{-\frac{ip^2}{2eE} - \frac{T^2 (p+\omega_p -m_\psi)^2}{2(2+ieET^2)}} \int_0^\infty dX X^{\nu_\phi^*} e^{-\beta X^2 -\gamma X} ~, 
\label{bigeq1}
\eeq
where
\beq
\beta = \frac{i}{2} +\frac{eET^2}{2+ieET^2} ~,
\eeq
\beq
\gamma =  \frac{2\sqrt{2} ip + \sqrt{2} eET^2 (\omega_p - m_\psi)} {\sqrt{eE}(2+ieET^2)} ~.
\eeq
The integral in the last equation can be expressed in terms of a parabolic cylinder function using Eq.~(3.462.1) of Ref.~\cite{GR},
\beq
\int_0^\infty dX X^{\nu_\phi^*}\, e^{-\beta X^2 -\gamma X} = \Gamma(1+\nu_\phi^*) (2\beta)^{-\frac{\nu_\phi^* +1}{2}} e^{\frac{\gamma^2}{8\beta}} D_{-(\nu_\phi^* +1)} \left(\frac{\gamma}{\sqrt{2\beta}}\right) .
\eeq

We are interested in the $p$-dependence of the amplitude.  
From Eqs. (\ref{eET}) and (\ref{eET2}), it follows that
$|\gamma/\sqrt{2\beta}|\gg |\lambda_\phi|^{1/2} \gg 1$.
Let us first consider the range $|\gamma/\sqrt{2\beta}|\gg |\lambda_\phi|$. Then, using the asymptotic form of the parabolic
cylinder function for $|Z|\gg \nu \sim |\lambda_\phi|$, we have:
\beq
e^{Z^2/4} D_\nu(Z) \propto Z^\nu .
\label{approxformula}
\eeq
Then it is easy to see that the dominant factor determining the $p$-dependence in this regime is the exponential factor in Eq.~(\ref{bigeq1}),
\beq
\exp\left[- \frac{T^2 (p+\omega_p -m_\psi)^2}{2(2+ieET^2)}\right] .
\eeq
Hence,
\beq
|{\cal A}_\chi(p;q=0)|^2 \propto \exp \left[-\frac{2\,(p+\omega_p -m_\psi)^2}{(eET)^2}\right].
\eeq
Strictly speaking, the asymptotic expansion (\ref{approxformula}) is not valid in the range
$|\lambda_\phi|\gg |\gamma/\sqrt{2\beta}|\gg |\lambda_\phi|^{1/2}$.
Nonetheless, by using a different expansion which is valid for
$|\lambda_\phi|\gg 1$, it can be checked that (F.10) is still valid in this regime \cite{GKT}.

\end{document}